%% file: main.tex
\documentclass[prd,aps,twocolumn,tightenlines,notitlepage,nofootinbib,preprintnumbers,letterpaper,superscriptaddress]{revtex4-2} 
\usepackage[normalem]{ulem}
\pdfoutput=1

\input{Packages}

\input{Commands}

\begin{document}

\begin{CJK*}{UTF8}{gbsn}
\title{Searching for sub-eV Sterile Neutrinos in Neutrino Telescopes}

\author{Emilse~Cabrera\orcidC{}}
\email{emilsecc@aluno.puc-rio.br}
\affiliation{Departamento de F\'isica, Pontif\'icia Universidade Cat\'olica do Rio de Janeiro,\\
Rua Marquês de São Vicente 225, Rio de Janeiro, Brazil}

\author{Miaochen Jin~(靳淼辰)\orcidD{}{}}
\email{miaochenjin@g.harvard.edu}
\affiliation{Department of Physics \& Laboratory for Particle Physics and Cosmology, Harvard University, Cambridge, MA 02138, USA}

\author{Carlos~A.~Arg{\"u}elles\orcidA{}}
\email{carguelles@fas.harvard.edu}
\affiliation{Department of Physics \& Laboratory for Particle Physics and Cosmology, Harvard University, Cambridge, MA 02138, USA}

\author{Arman~Esmaili\orcidB{}}
\email{arman@puc-rio.br}
\affiliation{Departamento de F\'isica, Pontif\'icia Universidade Cat\'olica do Rio de Janeiro,\\
Rua Marquês de São Vicente 225, Rio de Janeiro, Brazil}

\date{September 2025}

\begin{abstract}
With the forthcoming deployment of IceCube-Upgrade, unprecedented statistics of atmospheric neutrinos in the energy range (1-100) GeV will become available, providing a valuable opportunity to probe physics beyond the Standard Model in the neutrino sector. 
In this study, we calculate the sensitivity of the IceCube-Upgrade to sterile neutrinos with mass-squared splittings $\lesssim 1~{\rm eV}^2$. We demonstrate that, particularly due to the (1-10) GeV energy window, $\nu_\mu-\nu_s$ mixing angles as small as $\sim5^\circ$ can be probed by IceCube-Upgrade for all mass-squared splittings below $1~{\rm eV}^2$. Furthermore, we investigate the potential impact of a sterile neutrino state on the precision determination of standard atmospheric neutrino mixing parameters in the IceCube-Upgrade. 
\end{abstract}
\maketitle
\end{CJK*}

\section{Introduction\label{sec:intro}}

The search for sterile neutrinos has been actively pursued through numerous experiments and diverse methodologies for over two decades. Initially, sterile neutrinos with active-sterile mass-squared splittings of approximately $1~{\rm eV}^2$ were motivated by anomalies observed in the LSND~\cite{LSND:2001aii} and MiniBooNE~\cite{MiniBooNE:2013uba} experiments.
However, various other mass ranges have since garnered interest for different theoretical and observational reasons. 
Sterile neutrinos with heavier masses, ranging from keV up to around GeV, have been proposed in frameworks involving dynamical electroweak symmetry breaking~\cite{Appelquist:2002me} or within the neutrino Minimal Standard Model ($\nu$MSM)~\cite{Asaka:2005an,Asaka:2005pn}, where they could also serve as viable dark matter candidates. Collider and beam-dump experiments~\cite{Chun:2019nwi,Alekhin:2015byh}, as well as supernova observations~\cite{Fuller:2008erj,Rembiasz:2018lok,Mastrototaro:2019vug}, offer promising avenues for exploring these heavier sterile neutrino scenarios.

At the eV scale, sterile neutrinos not only provide potential resolutions to the aforementioned experimental anomalies but also play a role in facilitating r-process nucleosynthesis within supernovae~\cite{Caldwell:1999zk,Tamborra:2011is,Wu:2013gxa} and have profound implications for cosmology and the formation of large-scale structures~\cite{Dolgov:2002wy,Lesgourgues:2014zoa}. Numerous experimental strategies have been employed to probe eV-scale sterile neutrinos, including short-baseline neutrino experiments~\cite{LSND:2001aii,MiniBooNE:2013uba,MiniBooNE:2018Excess,MicroBooNE:2022Disappearance,MiniBooNE_SciBooNE:2011,NEOS:2017,DANSS:2018,DANSS:2022,DANSS:2023}, searches for averaged oscillation effects in long-baseline neutrino experiments~\cite{MINOS:2010fgd,MINOS:2011ysd,MINOS:2017cae,NOvA:2024imi,T2K:2019efw}, analysis of neutrinos emitted from supernovae bursts~\cite{Esmaili:2014gya,Choubey:2007ga,Warren:2014qza,Franarin:2017jnd}, and studies involving atmospheric neutrinos~\cite{SuperK:2015Sterile,ANTARES:2019JHEP,IceCube:2016PRL,IceCube:2017DeepCore,IceCube:2020PRL,IceCube:2020PRDLong,IceCube:2024DeepCore,IceCube:2024PRL,IceCube:2022Unstable}. For a review of these efforts see~\cite{Kopp:2013vaa,Collin:2016aqd,Diaz:2019fwt,Moulai:2019gpi,Hardin:2022muu,Boser:2019rta,Danilov:2022str,Acero:2022wqg,Dentler:2018sju}. In the lighter mass range, sterile neutrinos with active-sterile mass splitting $\sim10^{-5}~{\rm eV}^2$ have been proposed~\cite{deHolanda:2003tx,deHolanda:2010am} as a potential explanation for the absence of the expected upturn at low energies in the solar neutrino spectrum~\cite{Borexino:2017rsf}. In addition, mass splittings $\sim (10^{-5}-10^{-2})~{\rm eV}^2$ have been proposed~\cite{deGouvea:2022kma} as a solution to the tension between the T2K and NOvA data. Very light sterile neutrinos have been proposed to relax the cosmological bounds on neutrino masses~\cite{Ota:2024byu}. A complementary possibility is the quasi-Dirac scenario, in which each active state pairs with a nearly degenerate sterile partner; tiny splittings $\delta m^2 \sim 10^{-21}$–$10^{-11}\,\mathrm{eV}^2$ ~\cite{Wolfenstein:1981kw,Petcov:1982ya,Valle:1983dk,Doi:1983wu,Kobayashi:2000md}. This scenario is further motivated by recent arguments for Dirac neutrinos from Swampland conjectures and high-scale lepton-number breaking~\cite{Ooguri:2016pdq,Ibanez:2017kvh,Gonzalo:2021zsp,Casas:2024clw}. The resulting hyperfine splittings induce active–sterile oscillations over astronomical baselines, testable with solar~\cite{deGouvea:2009fp,Ansarifard:2022kvy}, supernova~\cite{DeGouvea:2020ang,Martinez-Soler:2021unz}, and high-energy astrophysical neutrinos~\cite{2000ApJS,2002ApJS,Beacom:2003eu,Keranen:2003xd,Esmaili:2009fk,Esmaili:2012ac,Joshipura:2013yba,Shoemaker:2015qul,Brdar:2018tce,Carloni:2022cqz,Fong:2024mqz,Dev:2024yrg,Carloni:2025dhv,MacDonald:2025spe}.

Searches for sterile neutrinos utilizing atmospheric neutrinos typically focus on resonance-enhanced active-to-sterile neutrino conversions induced by Earth’s matter effects, predominantly at energies greater than approximately $100~{\rm GeV}$~\cite{Nunokawa:2003ep, Esmaili:2012nz, Esmaili:2013cja, Esmaili:2013vza, Choubey:2007ji, Razzaque:2011ab, Razzaque:2012tp, Cabrera:2024rgi}; see also Refs.~\cite{Akhmedov:1988kd,Krastev:1989ix,Chizhov:1998ug,Chizhov:1999az,Akhmedov:1999va} for relevant Earth's matter effects discussions. Using this effect, the latest iteration of the IceCube high-energy muon-neutrino~\cite{IceCube:2024PRL} in the search for sterile neutrino has found a preference for light sterile neutrinos at the $2.2\sigma$ level.
In contrast, lower energy measurement using IceCube-DeepCore~\cite{IceCube:2017DeepCore} has failed to find signatures for light sterile neutrino for the comparable mass-square difference. In this work, we advocate for exploring sub-eV sterile neutrinos by atmospheric neutrinos at energies below $100~{\rm GeV}$, highlighting the significant sensitivity within the (1-10)~GeV energy range\footnote{For Earth's matter effect on sub-eV sterile neutrinos with energy $\lesssim$~GeV, see~\cite{Liao:2014ola}. For the effect of very light sterile neutrinos on the mass-ordering determination in long-baseline experiments, see~\cite{Chatterjee:2023qyr}.}. 

Our analysis is particularly timely given the forthcoming IceCube-Upgrade, which extends the IceCube detector’s sensitivity down to approximately $1~{\rm GeV}$, thus opening new avenues for detecting sterile neutrino signatures. The IceCube-Upgrade will consist of seven new strings deployed in the inner part of IceCube~\cite{Ishihara:2019aao}, where each string contains 100 optical modules. The IceCube-Upgrade optical modules are a combination of next-generation sensors with calibration devices~\cite{Rott:2025otm}. The IceCube-Upgrade is expected to significantly improve the detection efficiency and reconstruction in the sub-10 GeV energy region, while also reducing systematic uncertainties due to ice modeling. See Ref.~\cite{Dutta:2025hzs} for a recent study on the limitations of angular reconstruction in these experiments. In addition to the IceCube-Upgrade, the KM3NeT/ORCA detector~\cite{KM3NeT:2021ozk} is being deployed in the Mediterranean Sea. Using partial detector configuration, they have recently presented their first analysis of atmospheric neutrino oscillations~\cite{KM3NeT:2024ecf}. This detector is expected to be as competitive as the IceCube-Upgrade~\cite{Arguelles:2022hrt} in the determination of atmospheric oscillation parameters. Unfortunately, there is no detailed publicly available simulation for KM3NeT/ORCA, and thus we do not use it in our forecast and focus on the IceCube-Upgrade, where a detailed simulation is available.

The rest of this article is structured as follows: In Section~\ref{sec:pheno}, we briefly review the phenomenology of light sterile neutrinos and their atmospheric oscillation pattern. In Section~\ref{sec:analysis} we discuss the framework used in this work, including the simulation and statistical analysis. In Section~\ref{sec:results} we show the results of the analysis, including the impact of light sterile neutrinos on expected event rates and the sensitivity contours in the sterile neutrino parameter space. We also discuss the influence of sterile neutrinos on the standard neutrino oscillation picture. The conclusion is provided in Section~\ref{sec:conclusion}.

\section{Oscillation in the presence of sterile neutrinos}\label{sec:pheno}

In this section, we briefly describe the phenomenology of the $(3+1)$ sterile neutrino model, focusing particularly on the oscillation pattern of atmospheric neutrinos in the presence of a single, very light sterile neutrino state.
The inclusion of a fourth neutrino mass eigenstate, $m_4$, requires extending the conventional PMNS mixing matrix to a $4 \times 4$ unitary matrix, $U_{3+1}$, parameterized by six mixing angles and three CP-violating phases (excluding Majorana phases, which do not affect oscillation probabilities).
We adopt the following parameterization for $U_{3+1}$:
\begin{align}
    U_{3+1} = R^{34}(\theta_{34}) & R^{24}_{\delta_{24}}(\theta_{24}) R^{14}_{\delta_{14}} (\theta_{14})\times \\ & R^{23}(\theta_{23}) R^{13}_{\delta_{13}} (\theta_{13})R^{12}(\theta_{12})~,\nonumber
\end{align}
where $R^{ij}$ (with $i,j = 1,2,3,4$, adjacent indices $i<j$) denotes the rotation matrix in the $ij$-plane characterized by the angle $\theta_{ij}$.
The matrices $R^{ij}_{\delta_{ij}}$ involve rotations with non-adjacent $i$ and $j$ and include the CP-violating phase $\delta_{ij}$.
Notably, $\delta_{13}$ corresponds to the unique CP-violating phase present in the standard three-neutrino scheme.

The oscillation probabilities for atmospheric neutrinos can be derived by solving the evolution equation
\begin{equation}\label{eq:sch}
    i\frac{{\rm d}\nu_\alpha}{{\rm d}r} = \mathcal{H}_{\alpha\beta}\nu_\beta~,
\end{equation}
where the effective Hamiltonian $\mathcal{H}$ is given by
\begin{equation}
    \mathcal{H} = \frac{1}{2E_\nu} \left[ U_{3+1} M^2 U_{3+1}^\dagger + A(r)\right]~,
\end{equation}
with $M^2$ being the diagonal matrix containing the mass-squared differences $\Delta m_{ij}^2\equiv m_i^2 - m_j^2$, and $E_\nu$ representing the neutrino energy. The term $A(r)$ accounts for matter effects within Earth, explicitly dependent on the position $r$:
\begin{equation}
    A(r) = 2\sqrt{2}\, G_F E_\nu \,{\rm diag}\left(N_e(r),0,0,N_n(r)/2\right)~,
\end{equation}
where $G_F$ is the Fermi constant, and $N_e(r)$ and $N_n(r)$ denote the electron and neutron number densities of Earth, respectively. In our numerical analysis, we employ the \texttt{nuSQuIDS} package~\cite{nusquids}, which integrates the Preliminary Reference Earth Model (PREM)~\cite{PREM} to model Earth's density profile and solves Eq.~\eqref{eq:sch}. The oscillation probabilities are thus computed numerically as functions of neutrino energy $E_\nu$ and zenith angle $\theta_z$. In the following, we briefly review the key features of the oscillation patterns arising from the introduction of the sterile neutrino state.   

\begin{figure}[ht]
    \includegraphics[width=\textwidth]{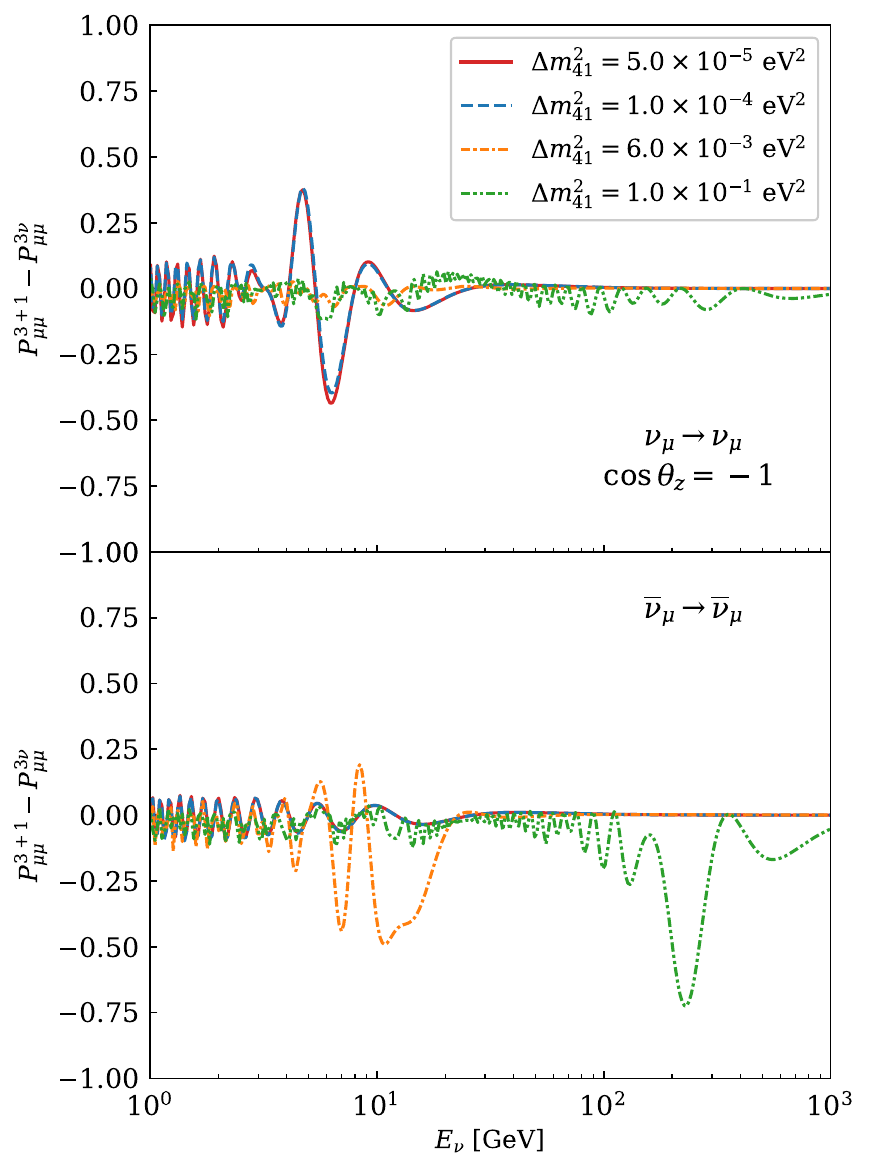}
    \caption[]{\textbf{\textit{Oscillation Probability Difference.}}
    Difference in the survival probability of core-crossing muon neutrinos (top panel) and muon antineutrinos (bottom panel) relative to the standard three-neutrino oscillation framework, for different sterile mass splittings in the sub-eV scale. 
    The standard parameters are set to their best-fit values~\cite{Esteban:2024eli}, while for the sterile sector we chose $\sin^2\theta_{24} = 0.03$, with all other active-sterile mixing angles set to zero.
    }
  \label{fig:oscillation-figure}
\end{figure}

To maintain computational feasibility, we focus our analysis on the parameter space spanned by $\theta_{24}$ and $\Delta m_{41}^2$, fixing $\theta_{34}=\theta_{14}=\delta_{24}=\delta_{14}=0$. This choice is further motivated by the observation that nonzero values of these parameters impose a stronger constraint on $\theta_{24}$~\cite{Esmaili:2013vza}. 
It is well-known that a nonzero value of $\theta_{24}$ induces MSW~\cite{Mikheyev:1985zog,Nunokawa:2003ep,Esmaili:2012nz,Esmaili:2013cja,Choubey:2007ji} and parametric~\cite{Liu:1997yb,Liu:1998nb} resonances in the $\bar{\nu}_\mu\to\bar{\nu}_s$ conversion channel (for $\Delta m_{41}^2>0$), occurring respectively at energies $E_\nu \sim 4\, {\rm TeV}\, (\Delta m_{41}^2/{\rm eV}^2)$ and $E_\nu \sim 2.3 \, {\rm TeV}\, (\Delta m_{41}^2/{\rm eV}^2)$, for trajectories passing the Earth's core ($\cos\theta_z=-1$). Given the IceCube-Upgrade's sensitivity in the energy range $1 \lesssim E_\nu/{\rm GeV}\lesssim 100$, these resonances become relevant and testable within the interval $\Delta m_{31}^2\leq\Delta m_{41}^2\lesssim 2\times10^{-2}~{\rm eV}^2$. For $\Delta m_{41}^2<\Delta m_{31}^2$ the resonances disappear from the $\bar{\nu}_\mu$ channel, mainly because the simple two-level oscillation picture breaks down; all four flavors mix simultaneously, distributing conversion probabilities across multiple channels. For $\Delta m_{41}^2\gtrsim 2\times10^{-2}~{\rm eV}^2$, the rapid oscillations in the tail of the resonances remain within IceCube-Upgrade's sensitivity window, providing another avenue for probing sterile neutrinos. These features are illustrated in the lower panel of Fig.~\ref{fig:oscillation-figure}, which shows modifications to the survival probability $P(\bar{\nu}_\mu\to\bar{\nu}_\mu)$ due to the presence of a sterile neutrino state along core-crossing paths, assuming $\sin^2\theta_{24}=0.03$. The pronounced resonances and rapid oscillation tails are distinctly visible for the orange dot-dashed ($\Delta m_{41}^2 = 6\times10^{-3}~{\rm eV}^2$) and green double-dot-dashed ($\Delta m_{41}^2 = 10^{-1}~{\rm eV}^2$) curves. In contrast, for smaller values such as $\Delta m_{41}^2 = 10^{-4}~{\rm eV}^2$ and $5\times10^{-5}~{\rm eV}^2$ (blue dashed and red solid curves, respectively), the resonances in the $\bar{\nu}_\mu$ channel become negligible. 

For $\Delta m_{41}^2\lesssim\Delta m_{31}^2$, which means $\Delta m_{43}^2 = \Delta m_{41}^2-\Delta m_{31}^2 <0$, the $\nu_\mu$ channel exhibits a resonance at $E_\nu\sim6$~GeV, for core-crossing trajectories, similar to the 1-3 resonance within the $\bar{\nu}_e-\bar{\nu}_{\mu,\tau}$ system that is governed by $\theta_{13}$, but now in the $\nu_\mu-\nu_s$ sector and governed by $\theta_{24}$. This resonance remains independent of the value $\Delta m_{41}^2$ as long as $m_4<m_3$, since $\Delta m_{43}^2 \to-\Delta m_{31}^2$ when $\Delta m_{41}^2\to0$. This behavior is clearly depicted in the upper panel of Fig.~\ref{fig:oscillation-figure}, where the blue dashed ($\Delta m_{41}^2 = 10^{-4}~{\rm eV}^2$) and red solid ($\Delta m_{41}^2 = 5\times10^{-5}~{\rm eV}^2$) curves prominently highlight this resonance.  

The combined effect of resonances on the survival probabilities of $\nu_\mu$ and $\bar{\nu}_\mu$ significantly modifies the zenith angle and energy distributions of atmospheric neutrinos, making IceCube-Upgrade an ideal detector for investigating sub-eV sterile neutrinos. In the following sections, we detail the analysis methodology and present the obtained results.

\section{Analysis Framework}\label{sec:analysis}

In this study, we analyze the atmospheric neutrino events observable at the IceCube Neutrino Observatory~\cite{Achterberg_2006}, an ice-Cherenkov-based neutrino telescope capable of detecting neutrinos with energies above $\mathcal{O}(10)$ GeV. Our work specifically focuses on the sensitivity projections for the upcoming IceCube-Upgrade~\cite{IceCube_Collaboration2020-md}, designed to lower the detection energy threshold down to $\mathcal{O}(1)$ GeV. We employ a publicly available Monte Carlo simulation dataset released by the IceCube Collaboration~\cite{IceCube_Collaboration2020-md}. To compute the weighted event rate from this Monte Carlo simulation, we utilize open-source packages and established models: atmospheric neutrino fluxes are calculated using the \texttt{NuFlux} package~\cite{nuflux}, while neutrino oscillation probabilities are obtained through the \texttt{nuSQuIDS} package~\cite{nusquids}, which incorporates the Preliminary Reference Earth Model (PREM)~\cite{PREM} for Earth's density profile. 

\begin{table}[ht!]
\vspace{0.1cm}
\begin{tabular}{l|c}
\textbf{Systematic Source} & $1\sigma$ range \\
\hline\hline
\multicolumn{2}{l}{\textbf{Flux Systematics}} \\ 
  \hline\hline
Flux normalization below 1 GeV & 25\% \\
Flux normalization above 1 GeV  & 15\% \\
Flux tilt     & 20\% \\
Upgoing flux normalization   & 20\% \\
Downgoing flux normalization    & 20\% \\
$\nu / \bar{\nu}$ ratio     & 2\% \\
Flavor ratio     & 5\% \\

\hline
\multicolumn{2}{l}{\textbf{Cross Section Systematics}} \\ 
  \hline\hline
NC/CC ratio     & 20\% \\
NC Hadron production    & 10\% \\
$\nu_\tau$ production     & 25\% \\
Axial mass    & 10\% \\
\hline
\end{tabular}
\caption{\textbf{\textit{Flux and cross section systematic uncertainties, and their $1\sigma$ ranges, used in this work.}}}
\label{table:Systematics}
\end{table}

By integrating these simulation tools, we generate event rates and their distributions corresponding to different sets of oscillation parameters. Additionally, we systematically account for uncertainties arising from neutrino flux and cross section models. The considered systematics uncertainties and their $1\sigma$ ranges are summarized in Table~\ref{table:Systematics}. Furthemore, detector-related systematic uncertainties, as described in the data release~\cite{IceCube_Collaboration2020-md}, are also incorporated. All systematic uncertainties are included as fluctuations in the event-per-bin ratios.

\begin{table}
\begin{tabular}{l|c|c|c}
Parameter & Nominal & Boundary & Points \\
\hline\hline
$\sin^2\theta_{23}$ &  $0.572$    & [0.305, 0.705] & 15 \\
$\sin^2\theta_{24}$ &  $0.0$     & [$10^{-3}$, 1] & 30  \\
$\Delta m^2_{31}$ [eV$^2$]  &  $2.5\times 10^{-3}$     & [$2 \times 10^{-3}$, $3 \times 10^{-3}$] & 15 \\
$\Delta m^2_{41}$ [eV$^2$]  &  $0.0$    & [$10^{-5}$, 1] & 50 \\
Ordering            & Normal  & fixed & 1
\end{tabular}
\caption{\textbf{\textit{Summary of nominal values and boundary values in $(3+1)$ oscillation parameter space. All other parameters are fixed.}}}
\label{table:OscPar}
\end{table}

We use an array of $337,500$ points in the parameter space of oscillation parameters, summarized in Table~\ref{table:OscPar}. For each point, we compute the $\chi^2$ function, that is the log-likelihood test statistics for a Poisson distribution for events per bin, and across all the bins, and minimize by marginalizing over the systematics. The $\chi^2$ function is given by
\begin{widetext}
\begin{equation}
    \chi^2 = 2 \sum_{\textrm{bin} \, {i}} \left[ N^\textrm{mod}_i \left(1 + \sum_{\textrm{sys} \, \eta_j} f_i(\eta_j) \right) - N^{\textrm{dat}}_i + N^\textrm{dat}_i \cdot \log\left( \frac{N^\textrm{dat}_i}{\tilde{N}^\textrm{mod}_i(\eta_j)} \right) \right] + \sum_{\textrm{sys}\,\eta_j} \left( \frac{\bar{\eta}_j - \eta_j}{\sigma_j}\right)^2~,
    \label{eq:chi2}
\end{equation}
\end{widetext}
Where $N_i^{\rm mod}$ is the binned Asimov expected event rate (summing over tracks and cascades) assuming the standard 3-flavor oscillations; and $N_i^{\rm dat}$ are the MC simulation binned event rates at various sets of sterile mixing parameters. The first term in Eq.~\eqref{eq:chi2} is the log-likelihood test statistics and the second term is the penalty term for the nuisance parameter, where each nuisance parameter is assumed to have an independent Gaussian prior. The factor $f_i(\eta_j)$ is the fractional event rate change in bin $i$ due to the $j$-th systematics $\eta_j$, and we conveniently denote the systematic-modified model prediction in each bin as $\tilde{N}^\textrm{mod}_i(\eta_j)$. The $\bar{\eta}_j$ is the nominal value of the $j$-th systematic uncertainty, and $\sigma_j$ is the $1\sigma$ range. The minimization is performed using the analytical form of the Jacobian ${\rm d}\chi^2 / {\rm d}\eta_j$ which can be easily derived from~Eq.\eqref{eq:chi2}. The minimization procedure is similar to that in previous works~\cite{Arguelles:2022hrt}. Afterwards, we profile over the standard oscillation parameters to obtain the $\chi^2_\textrm{min}$ for each pair of $(\theta_{24}, \, \Delta m^2_{41})$ values, which can be interpolated to obtain the final sensitivity contours in the parameter space.

\begin{figure*}[ht]
    \includegraphics[width=\textwidth]{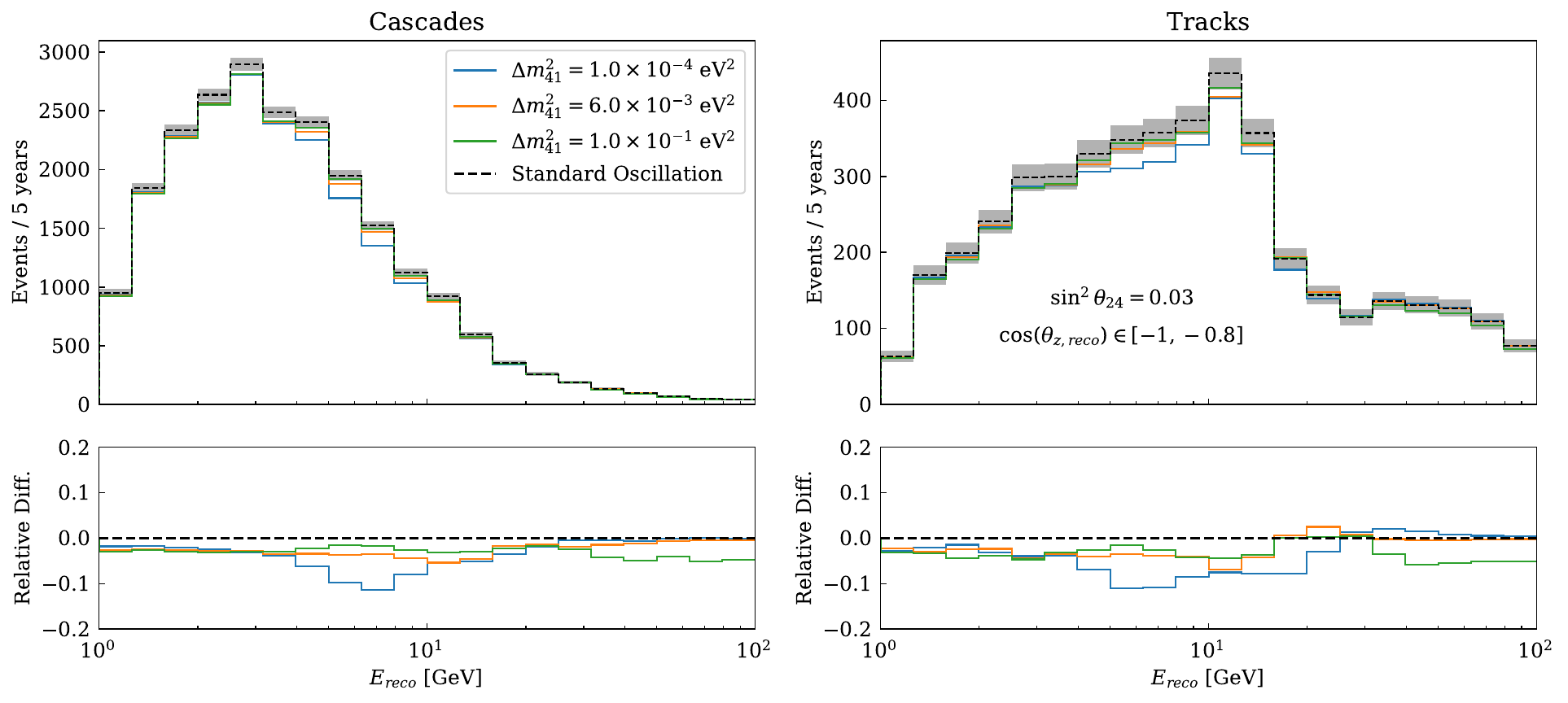}
    \caption[]{\textbf{\textit{Events Energy Distribution.}}
    (Top panels) Event distribution for selected values of the sterile mass splitting, matching the values in Fig.~\ref{fig:oscillation-figure}, as a function of the reconstructed neutrino energy, $E_{\rm reco}$, for $\cos (\theta_{z,{\rm reco}})\in [-1,-0.8]$. The band around the event distribution of $3\nu$ oscillation corresponds to the $\pm1\sigma$ statistical error. (Bottom panels) The relative difference in number of events with respect to standard oscillation (null hypothesis). The left and right panels correspond to cascade and track events, respectively. We choose $\sin^2\theta_{24}=0.03$.
    }
  \label{fig:dist1D}
\end{figure*}

\begin{figure}[ht]
    \includegraphics[width=\textwidth]{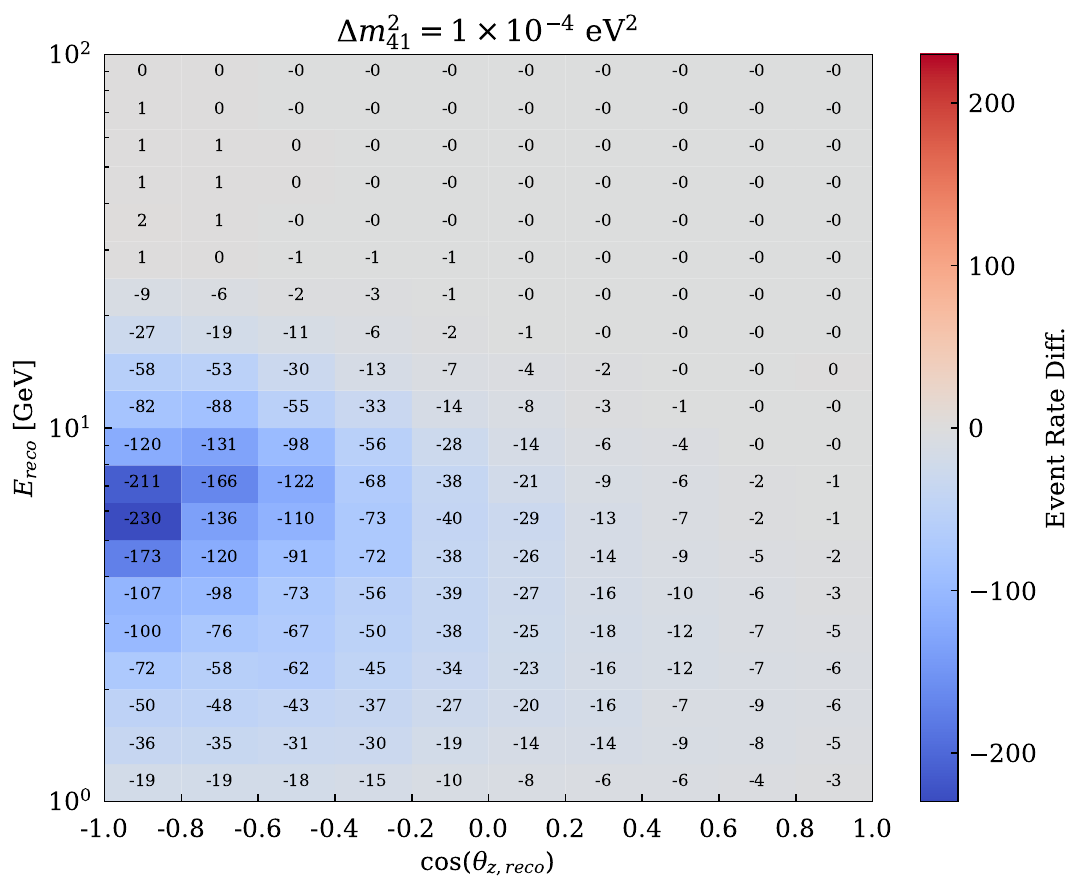}\\

 \includegraphics[width=\textwidth]{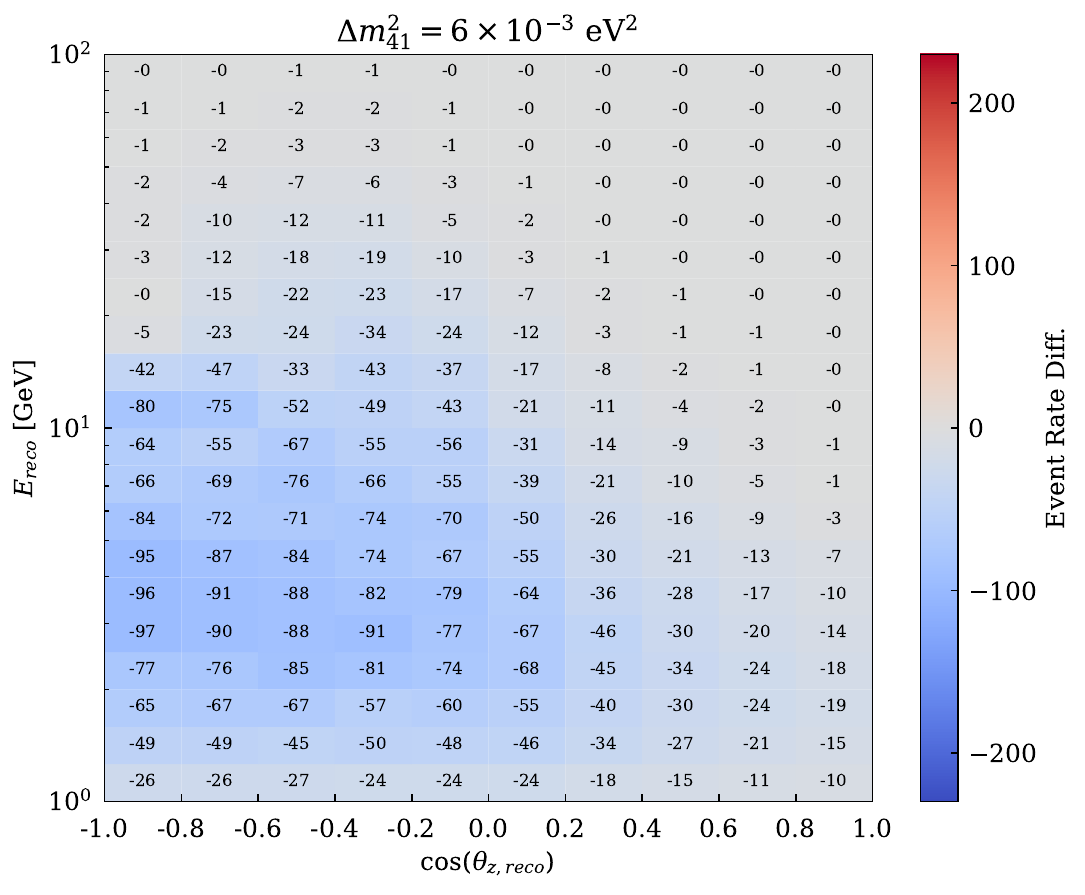}
    \caption[]{\textbf{\textit{2D Event Distribution.}}
    Impact of the sterile neutrino, with mass splittings $\Delta m_{41}^2 = 10^{-4}$ eV$^2$ (top panel) and $\Delta m_{41}^2 =6\times 10^{-3}$ eV$^2$ (bottom panel), and mixing $\sin^2\theta_{24} = 0.03$, on the morphology-summed event distribution in two reconstructed observables: $E_{\rm reco}$ and $\cos (\theta_{z,{\rm reco}})$. The numbers are the difference in number of events in the $(3+1)$ and standard scenarios.}
    
  \label{fig:dist2D}
\end{figure}

\section{Results}\label{sec:results}

In this section, we present the results of two analyses: the sensitivity projection of IceCube-Upgrade to the sterile parameter space of $(\sin^2\theta_{24},\Delta m_{41}^2)$, and the effect of a hypothetical sterile neutrino state on the determination of standard mixing parameters $(\sin^2\theta_{23},\Delta m_{31}^2)$.

Fig.~\ref{fig:dist1D} shows the impact of sterile neutrino for three distinct values of $\Delta m_{41}^2$ (matching those used in Fig.~\ref{fig:oscillation-figure}) on the energy distribution of events within the zenith bin $\cos(\theta_{z,{\rm reco}})\in [-1,-0.8]$ (where $\theta_{z,{\rm reco}}$ is the reconstructed zenith angle), assuming $\sin^2\theta_{24} = 0.01$. The upper panels depict cascade (left) and track (right) event distributions, with the shaded band indicating the $1\sigma$ statistical uncertainty. In the energy range $(1-100)$~GeV, cascades dominate partly because track events at low energies might be so short to be clearly distinguished from cascade-like events. This, combined with improvements in light sensor response from newly developed optical modules and deployed calibration devices, enhances cascade event reconstruction. Despite the higher statistics of cascades, track events remain crucial due to their superior angular resolution and because muon neutrinos dominate in the atmospheric flux, making them particularly sensitive probes of muon neutrino disappearance. Both event topologies exhibit notable distortions near the energy range of the resonances, particularly evident at $E_{\rm reco}\simeq 6$~GeV for $\Delta m_{41}^2\leq\Delta m_{31}^2$, exemplified by the blue curve corresponding to $\Delta m_{41}^2 = 10^{-4}$ eV$^2$. These distortions are highlighted in the lower panels, which display the relative difference in event numbers compared to the standard three-neutrino oscillation scenario. The pronounced deficit in the number of events at energies $\lesssim10$~GeV for $\Delta m_{41}^2\leq\Delta m_{31}^2$, arises from the resonance that occurs in the neutrino channel. Since the cross sections of neutrino interaction are approximately two times larger than those of antineutrinos, neutrino-induced events are predominant in the observable signal.

To illustrate the zenith-angle distortion, Fig.~\ref{fig:dist2D} shows the difference in event rate between sterile and standard scenarios, combining both cascade and track events as a two-dimensional distribution in reconstructed energy and zenith angle. We adopt $\sin^2\theta_{24}=0.01$, and $\Delta m_{41}^2=10^{-4}~{\rm eV}^2$ in the top panel and $\Delta m_{41}^2=6\times10^{-3}~{\rm eV}^2$ in the bottom panel. The top panel, representing $\Delta m_{41}^2\leq\Delta m_{31}^2$, clearly displays the suppression of events at $E_{\rm reco}\lesssim 10$~GeV and $\cos (\theta_{z,{\rm reco}})\sim [-1,0]$, promising the sensitivity of IceCube-Upgrade to very light sterile neutrinos through matter-enhanced disappearance effects in the neutrino channel. A sterile neutrino with mass splitting $\Delta m_{41}^2\geq\Delta m_{31}^2$ induces resonance in the antineutrino channel. For the case $\Delta m_{41}^2=6\times10^{-3}~{\rm eV}^2$ in the lower panel of Fig.~\ref{fig:dist2D}, the suppression occurs primarily at $E_{\rm reco}\simeq10$~GeV and $\cos (\theta_{z,{\rm reco}})\sim [-1,0]$. For higher mass splittings $\Delta m_{41}^2 \gtrsim10^{-2}~{\rm eV}^2$, the resonance moves beyond IceCube-Upgrade's sensitivity energy range; however, rapid oscillations persist at the resonance tail, producing observable distortions. This higher mass-splitting regime has been extensively explored in previous IceCube analyses~\cite{IceCube:2016PRL, IceCube:2020PRL, IceCube:2024PRL, IceCubeCollaboration:2024dxk}.

\begin{figure}[ht]
    \includegraphics[width=\textwidth]{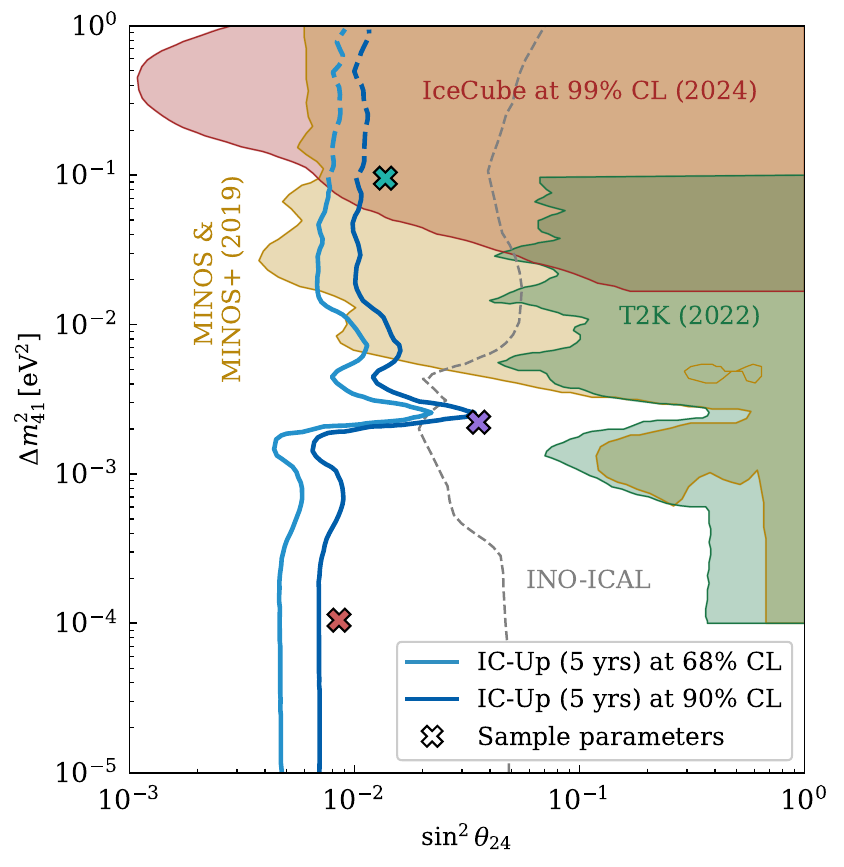}
    \caption[]{\textbf{\textit{Bounds on the Sterile Neutrino Parameter Space.}}
    Expected sensitivity at 68$\%$ CL (blue line) and at 90$\%$ CL (dark blue line) in the ($\sin^2 \theta_{24}, \Delta m_{41}^2$) plane, from 5 years of IceCube-Upgrade operation. The segments of exclusion curves depicted by the dashed line, corresponding to $\Delta m_{41}^2 \gtrsim 0.1 \; \rm{eV}^2$, are apt to changes by a more detailed analysis taking into account the distribution of neutrino production altitudes and experimental uncertainties. The shaded regions indicate the bounds derived by MINOS and MINOS+~\cite{MINOS:2017cae}, T2K~\cite{deGouvea:2022kma} at 90\% CL and IceCube~\cite{IceCube:2024PRL} at 99\% CL. The gray dashed curve shows the projected sensitivity of INO-ICAL~\cite{Thakore:2018lgn} at 90\% CL. The crosses represent sample parameter values used to study their impact on the standard oscillation parameters in Fig.~\ref{fig:scan-2}.
    }
  \label{fig:scan-1}
\end{figure}

The sensitivity of IceCube-Upgrade, after five years of operation, in the $(\sin^2\theta_{24},\Delta m_{41}^2)$ plane is presented in Fig.~\ref{fig:scan-1}. The blue and dark blue curves correspond to 68\% and 90\% CL exclusion limits, respectively. For reference, we include existing constraints from IceCube~\cite{IceCube:2024PRL}, MINOS/MINOS+~\cite{MINOS:2017cae}, and T2K~\cite{deGouvea:2022kma}, by the colored shaded regions. Additionally, the projected sensitivity of INO-ICAL~\cite{Thakore:2018lgn} for 500 kt-yr and at 90\% CL is shown by the dashed gray curve. The IceCube limit is at 99\% C.L. and shows the excluded region (the reported 90\% C.L. region in~\cite{IceCube:2024PRL} is a closed contour). IceCube also reported the DeepCore results~\cite{IceCube:2024dlz} which focuses on sterile mass squared splittings $\gtrsim 1~ \mathrm{eV}^2$, so we do not include it here. For $\Delta m_{41}^2\lesssim10^{-2}~{\rm eV}^2$, the IceCube-Upgrade is expected to improve the existing bounds from MINOS and T2K on $\sin^2\theta_{24}$ by nearly two orders of magnitude. This significant enhancement arises because, in long-baseline experiments such as MINOS and T2K, sterile neutrino effects manifest as subtle modulations of the neutrino energy spectra over the standard $3\nu$ oscillation pattern, with negligible matter effects. Compared to INO-ICAL, the IceCube-Upgrade demonstrates superior sensitivity by approximately one order of magnitude. For $\Delta m_{41}^2\gtrsim0.1~{\rm eV}^2$, the entire IceCube detector offers enhanced performance due to the larger statistics at higher energy, where the resonance in active-sterile conversion occurs. A comment on the range of $0.1 \lesssim \Delta m_{41}^2/{\rm eV}^2 \lesssim 1$ is in order. For this range of $\Delta m_{41}^2$ values, the vacuum oscillation length is $\sim (25-250)$~km at $E_\nu\sim10$~GeV, leading to oscillations for down-going and horizontal neutrinos (see Appendix~\ref{appx} for 2D event distributions of tracks and cascades for various $\Delta m_{41}^2$ values, as well as the corresponding oscillograms). Consequently, for this range of $\Delta m_{41}^2$ values, the altitude distribution of atmospheric neutrino production points should be taken into account consistently. In addition, special treatment of the atmospheric muon background is inevitable. Since these effects are beyond the scope of this work and our focus is on the very light sterile neutrino masses, the exclusion curves in Figure~\ref{fig:scan-1} for this region are shown by the dashed line, indicating that it is prone to changes.

For $\Delta m_{41}^2 \leq \Delta m_{31}^2$, the exclusion contours become nearly vertical, because the resonance in muon neutrino conversion occurs at similar energies across this parameter range (as illustrated in Fig.~\ref{fig:dist1D}). Although Fig.~\ref{fig:scan-1} displays results down to $\Delta m_{41}^2=10^{-5}~{\rm eV}^2$, these vertical exclusion contours continuously extend toward lower values of $\Delta m_{41}^2$, theoretically down to zero. We restrict our analysis to $\Delta m_{41}^2\geq0$ to avoid scenarios where all active neutrino states become heavier, strongly conflicting with cosmological constraints on the sum of neutrino masses (however, see~\cite{Ota:2024byu}. 

\begin{figure}[ht]
    \includegraphics[width=\textwidth]{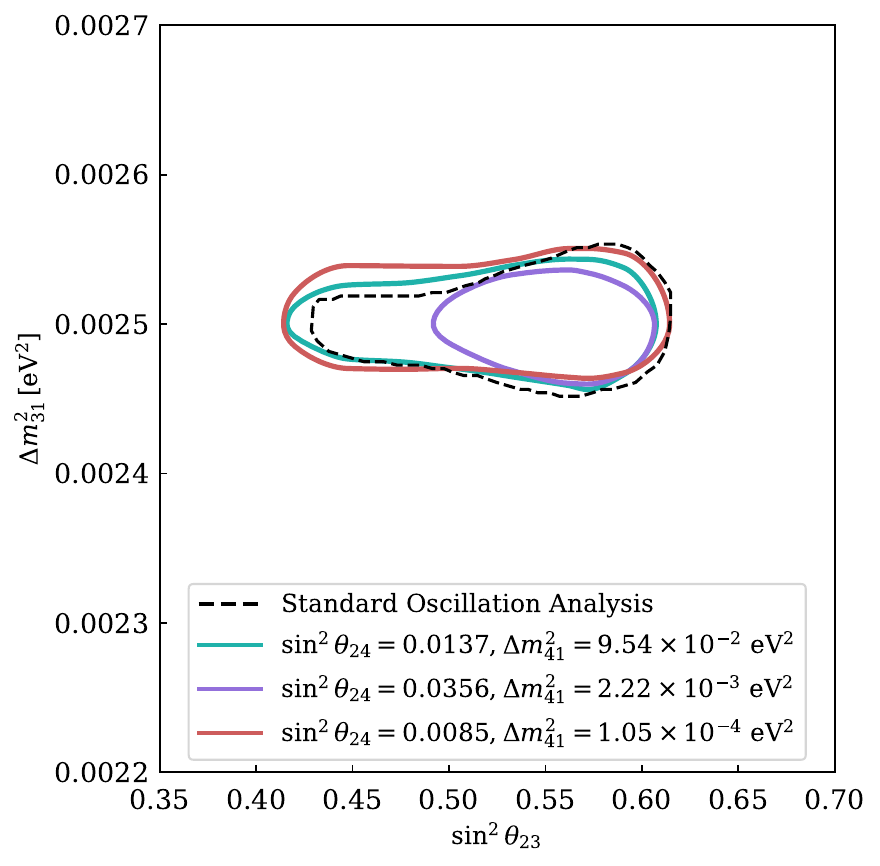}
    \caption[]{\textbf{\textit{Impact of Sterile Neutrino on the Determination of Atmospheric Oscillation Parameters.}}
        The 90$\%$ CL regions in the $(\sin^2 \theta_{23}, \Delta m^2_{31})$ plane, for three sterile neutrino scenarios, corresponding to the crosses in Fig.~\ref{fig:scan-1}. The dashed black line shows the result from the standard oscillation analysis without a sterile neutrino~\cite{Arguelles:2022hrt}.
    }
  \label{fig:scan-2}
\end{figure}

If a sterile neutrino with mass splitting $\Delta m_{41}^2\lesssim0.1~{\rm eV}^2$ and mixing angle $\sin^2\theta_{24}\gtrsim10^{-2}$ exists, IceCube-Upgrade is well-positioned to provide evidence of its presence. However, the existence of a sterile state with mass splitting and mixing values close to the exclusion curves shown in Fig.~\ref{fig:scan-1}, could potentially affect the determination of atmospheric neutrino parameters $\Delta m_{31}^2$ and $\sin^2\theta_{23}$. To evaluate this impact, we select three representative points near the sensitivity boundary, marked by the crosses in Fig.~\ref{fig:scan-1}, and investigate their influence on the determination of atmospheric parameter. Fig.~\ref{fig:scan-2} displays the contours in the $(\sin^2\theta_{23},\Delta m_{31}^2)$ plane for these points, compared to the reference contour without a sterile state (dashed black curve). While the influence is minimal for $\Delta m_{41}^2\simeq10^{-1}~{\rm eV}^2$ and $10^{-4}~{\rm eV}^2$, a significant distortion occurs at $\Delta m_{41}^2 \simeq \Delta m_{31}^2$. Specifically, in this degenerate region (shown by the violet curve in Fig.~\ref{fig:scan-2}), the active-sterile neutrino oscillation partially mimics the standard atmospheric neutrino oscillation pattern, artificially tightening the constraint on $\sin^2\theta_{23}$.

\section{Conclusions}\label{sec:conclusion}

With the upcoming deployment of IceCube-Upgrade, a low-energy extension of the IceCube detector, unprecedented statistics of atmospheric neutrino events in the energy range $(1-100)$~GeV will soon become available. The particularly valuable lower energy range of $(1-10)$~GeV—historically less explored—offers substantial potential for addressing key questions within and beyond the Standard Model, including neutrino mass hierarchy determination and the existence of sterile neutrino states. In this study, we investigated the capability of IceCube-Upgrade to probe the $(3+1)$ sterile neutrino scenario, featuring one additional sterile neutrino state beyond the Standard Model with a mass-squared splitting of $\lesssim 1~{\rm eV}^2$ relative to the lightest active neutrino. Our analysis was specifically restricted to mixing between mass eigenstates 2 and 4, parameterized by the angle $\theta_{24}$.

The observable signature of sterile neutrinos in atmospheric neutrino oscillations varies with the mass-squared splitting $\Delta m_{41}^2$. For $\Delta m_{41}^2 \geq \Delta m_{31}^2$, MSW and parametric resonances arise in the $\bar{\nu}_\mu \to \bar{\nu}_s$ channel (assuming normal hierarchy among active neutrinos and $\Delta m_{41}^2 \geq 0$), inducing distinctive zenith angle and energy-dependent distortions extensively studied in prior literature. In contrast, for $\Delta m_{41}^2 \leq \Delta m_{31}^2$, the resonant conversion of $\nu_\mu \to \nu_s$ provides sensitivity to arbitrarily small values of $\Delta m_{41}^2$, effectively down to zero.

Using publicly available Monte Carlo simulations for IceCube-Upgrade, our results indicate that, for mass splittings $\Delta m_{41}^2 \lesssim 10^{-2}~{\rm eV}^2$, the constraints on $\sin^2\theta_{24}$ can be enhanced by approximately two orders of magnitude after five years of data collection. This substantial improvement primarily stems from the enhanced statistics and sensitivity to atmospheric neutrinos in the $(1-10)$~GeV energy range. Furthermore, we examined the influence of a sterile neutrino with mixing near $\sin^2\theta_{24} \simeq 10^{-2}$, corresponding to the projected sensitivity limit of IceCube-Upgrade, on the determination of standard atmospheric neutrino parameters $\Delta m_{31}^2$ and $\sin^2\theta_{23}$. Notably, in the specific case of $\Delta m_{41}^2 \simeq \Delta m_{31}^2$, we demonstrated that sterile-induced degeneracy could lead to artificially tightened constraints on $\sin^2\theta_{23}$ derived from IceCube-Upgrade data. Comparison of these constraints with results from long-baseline accelerator experiments offers an indirect approach for testing this scenario.

\section*{Code Availability}
All relevant code and data can be found in \href{https://github.com/MiaochenJin/NeutrinoTelescopeOscillationAnalysis}{\textcolor{blue}{this github}}. For all other questions and reasonable requests, please directly contact the authors.

\acknowledgements{}
A.~E. thanks FAPERJ for partial financial support through the grant SEI-260003/005924/2024. E.~C. thanks the support from CNPq scholarship No. 140121/2022-6 and CAPES-PDSE scholarship No. 88881.982354/2024-01. E.~C. also thanks the Laboratory for Particle Physics and Cosmology for its hospitality, where part of this work was developed.
CAA and MJ are supported by the Faculty of Arts and Sciences of Harvard University, and are supported for a part of this work by the National Science Foundation (NSF). CAA is supported by the NSF AI Institute for Artificial Intelligence and Fundamental Interactions,
the Research Corporation for Science Advancement, and the David \& Lucile Packard Foundation.

\bibliography{sterile-light.bib}

\onecolumngrid
\newpage
\appendix

\section{Detailed Event Distributions and Oscillograms}\label{appx}

In this appendix, we show the oscillograms and the corresponding 2D event distributions, separately for tracks and cascades, for various values of $\Delta m_{41}^2$. For very light sterile neutrino masses, the sensitivity stems mainly from the up-going track events at $\lesssim10$~GeV, corresponding to $L/E_\nu \gtrsim 1000\, [\rm{km} / \rm{GeV}]$; while, for $\Delta m_{41}^2$ in the range of $(0.1-1)~\rm{eV}^2$, the sensitivity originates mainly from the down-going cascade events at $\gtrsim10$~GeV, corresponding to vacuum oscillation with $L/E_\nu \lesssim 250\, [\rm{km} / \rm{GeV}]$.

\begin{figure}[ht]
    \centering
    \includegraphics[width=0.4\linewidth]{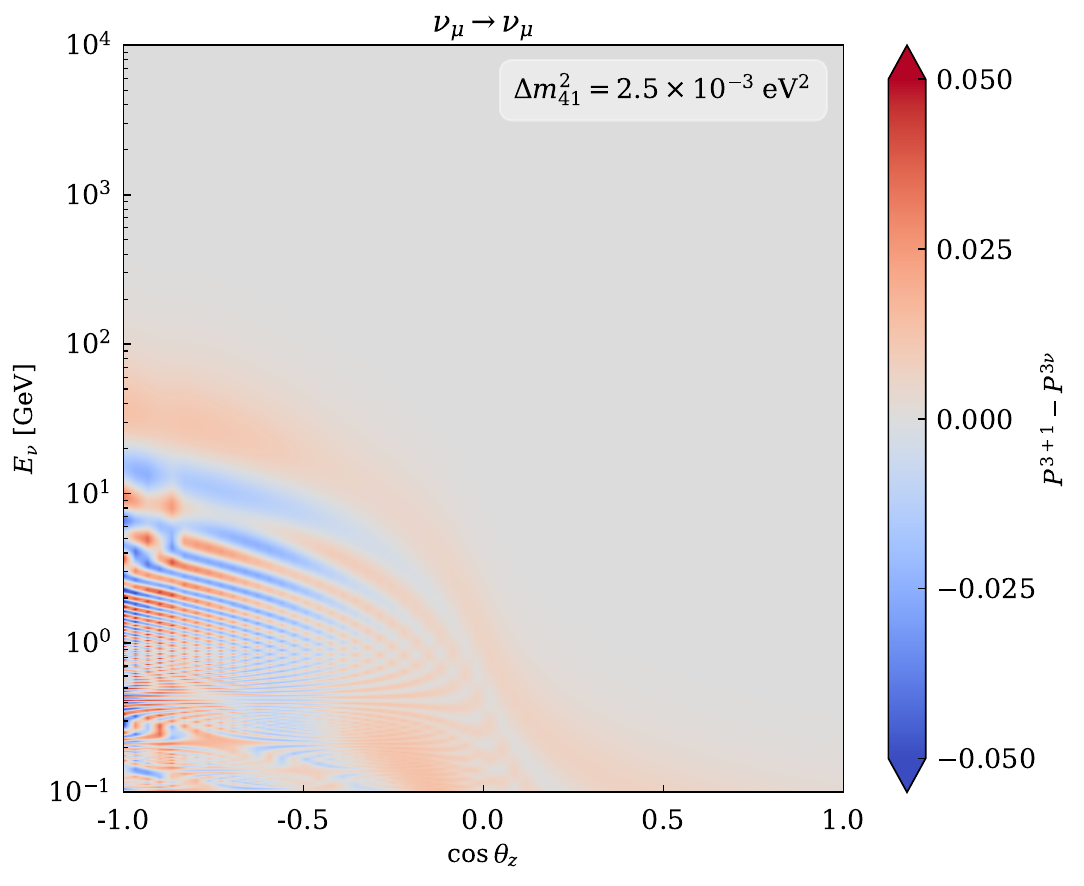}
    \includegraphics[width=0.4\linewidth]{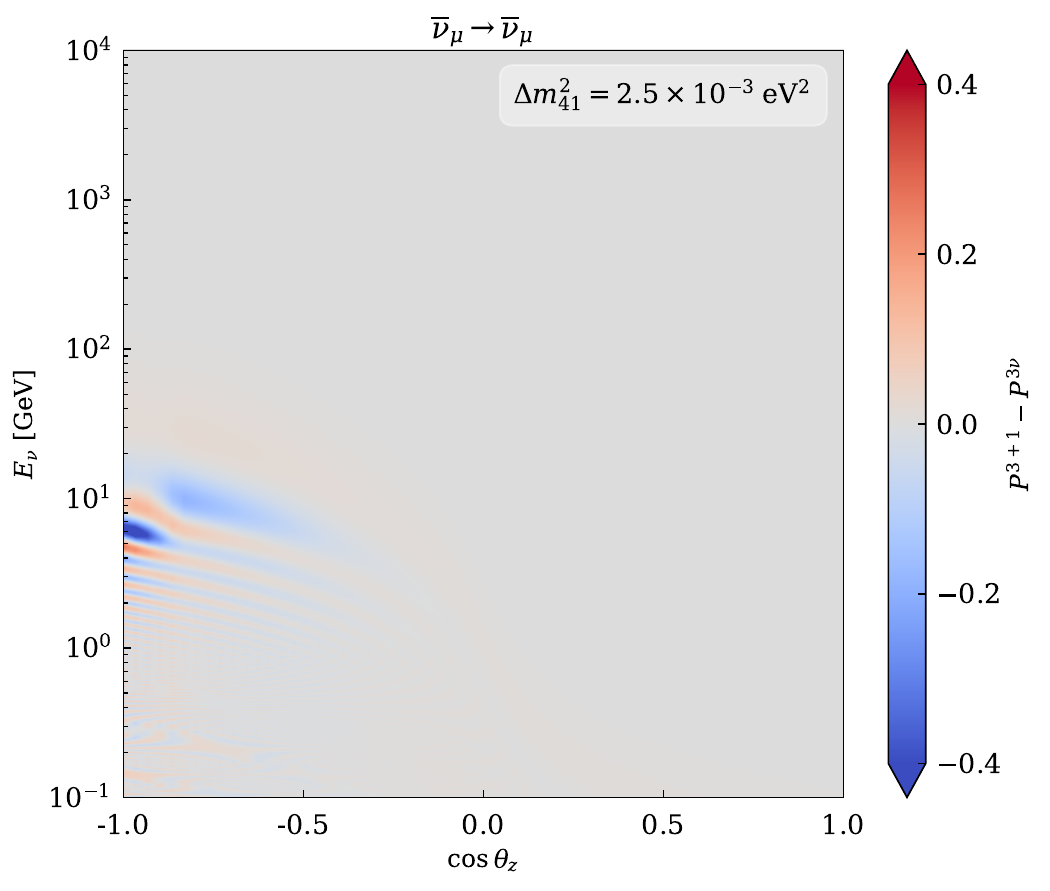}\\
    \includegraphics[width=0.4\linewidth]{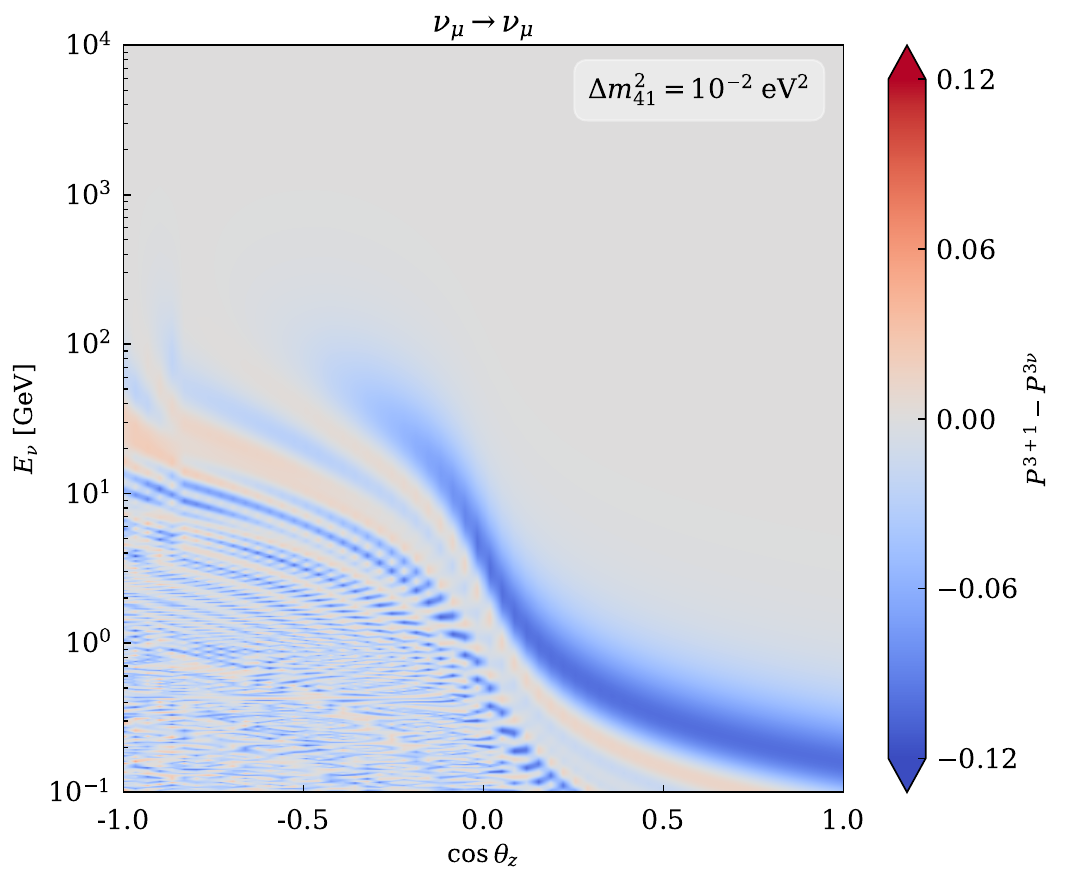}
    \includegraphics[width=0.4\linewidth]{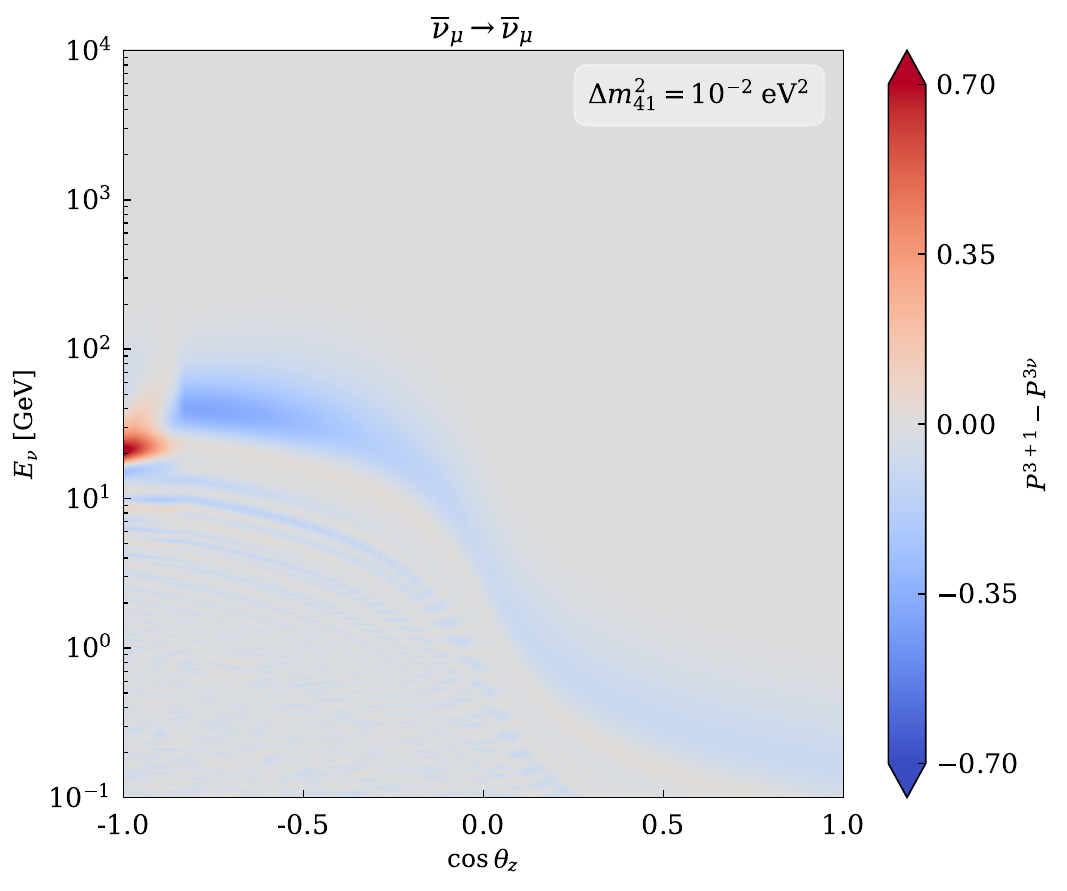}
    \caption{Oscillograms of the difference between the $3+1$ and $3\nu$ survival probabilities of $\nu_\mu$ (left) and $\bar{\nu}_\mu$ (right), for two representative values of the active-sterile mass splitting, $\Delta m_{41}^2 = 2.5\times10^{-3}$ and $ 10^{-2}$ eV$^2$.
    }    
    \label{fig:oscillograms1}
\end{figure}

\begin{figure}[ht]
    \centering
    \includegraphics[width=0.4\linewidth]{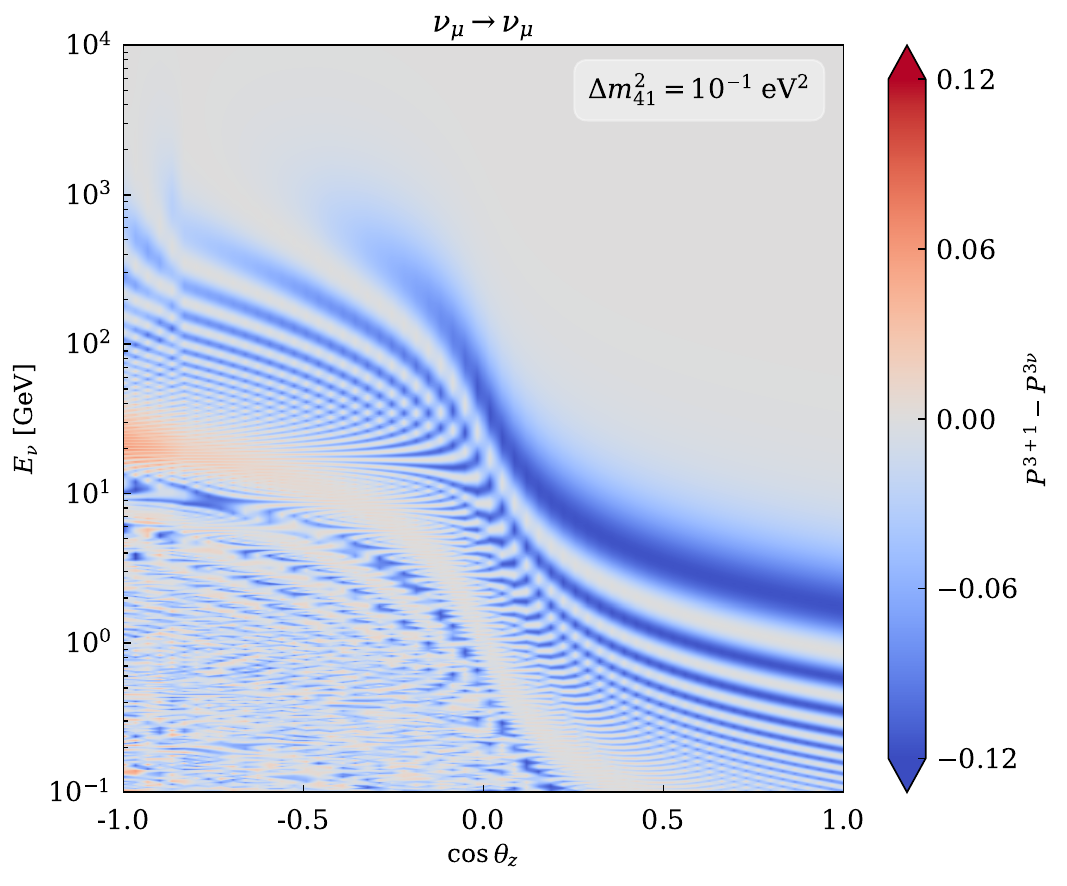}
    \includegraphics[width=0.4\linewidth]{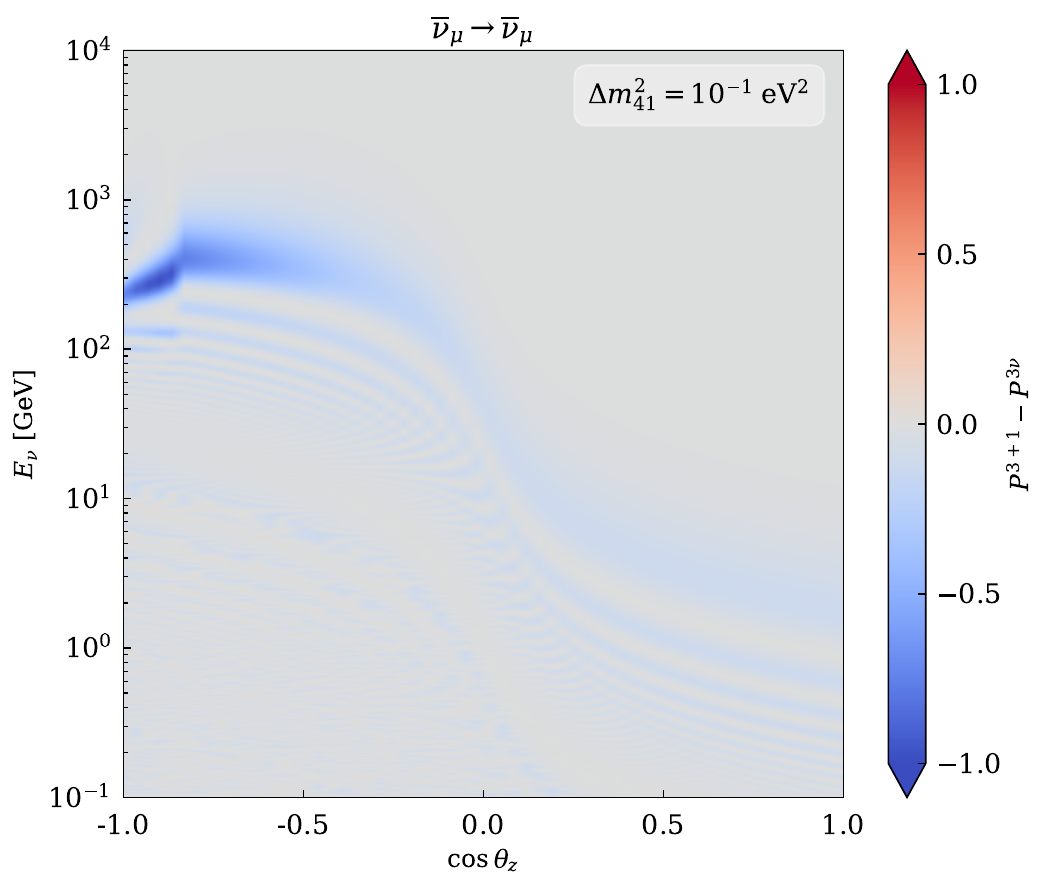}\\
    \includegraphics[width=0.4\linewidth]{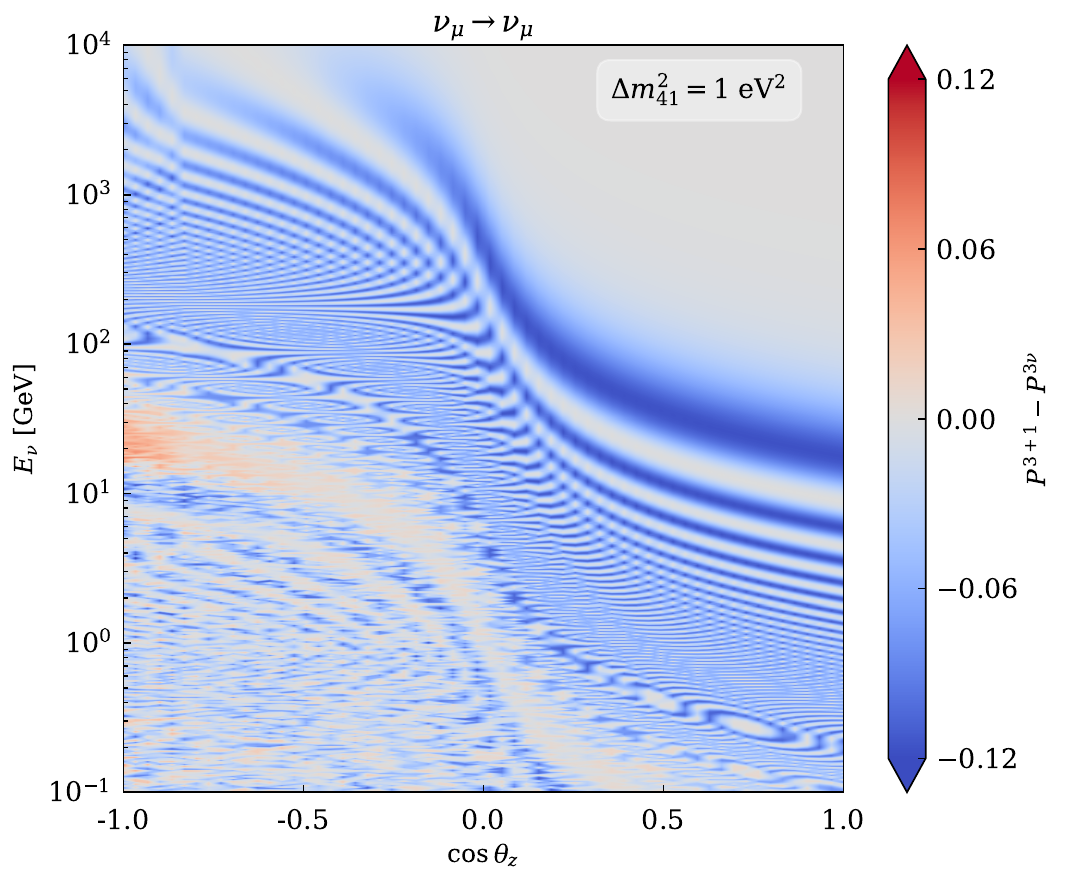}
    \includegraphics[width=0.4\linewidth]{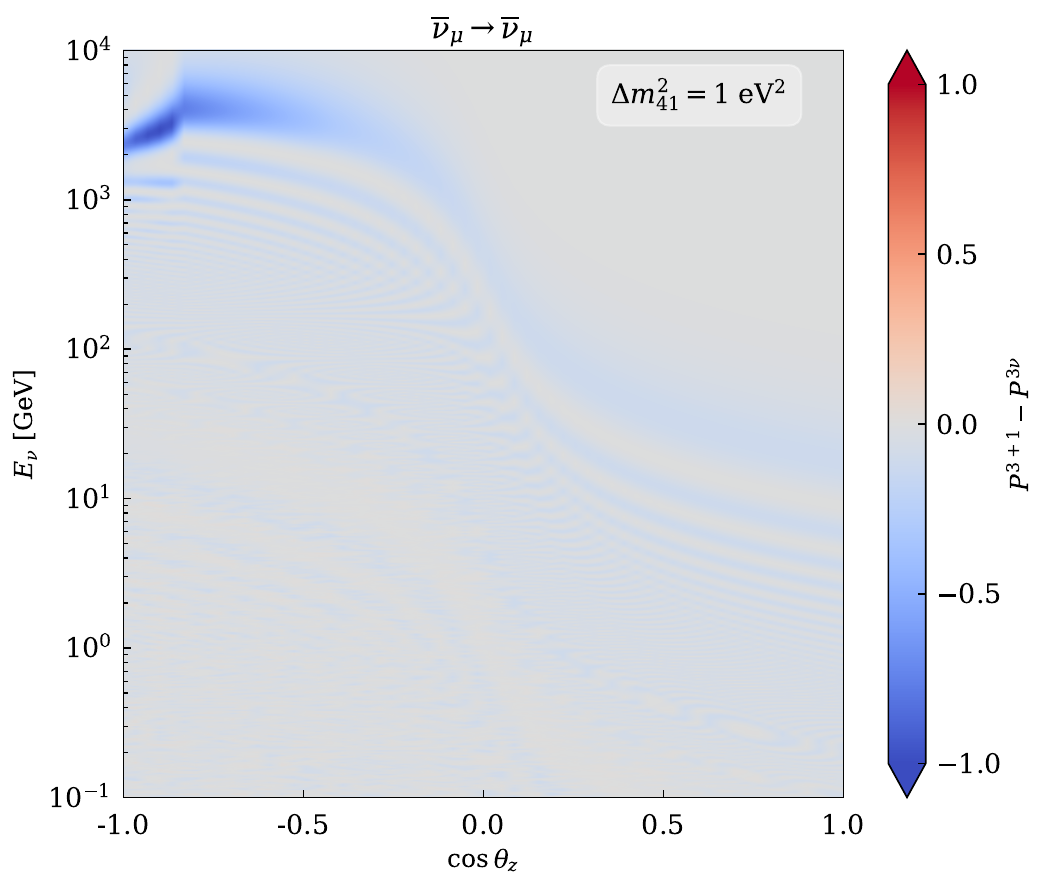}
    \caption{The same as Fig.~\ref{fig:oscillograms1}, but for $\Delta m_{41}^2 = 10^{-1}$ and 1 eV$^2$.}    
    \label{fig:oscillograms2}
\end{figure}

\begin{figure}[ht]
    \centering
    \includegraphics[width=0.4\linewidth]{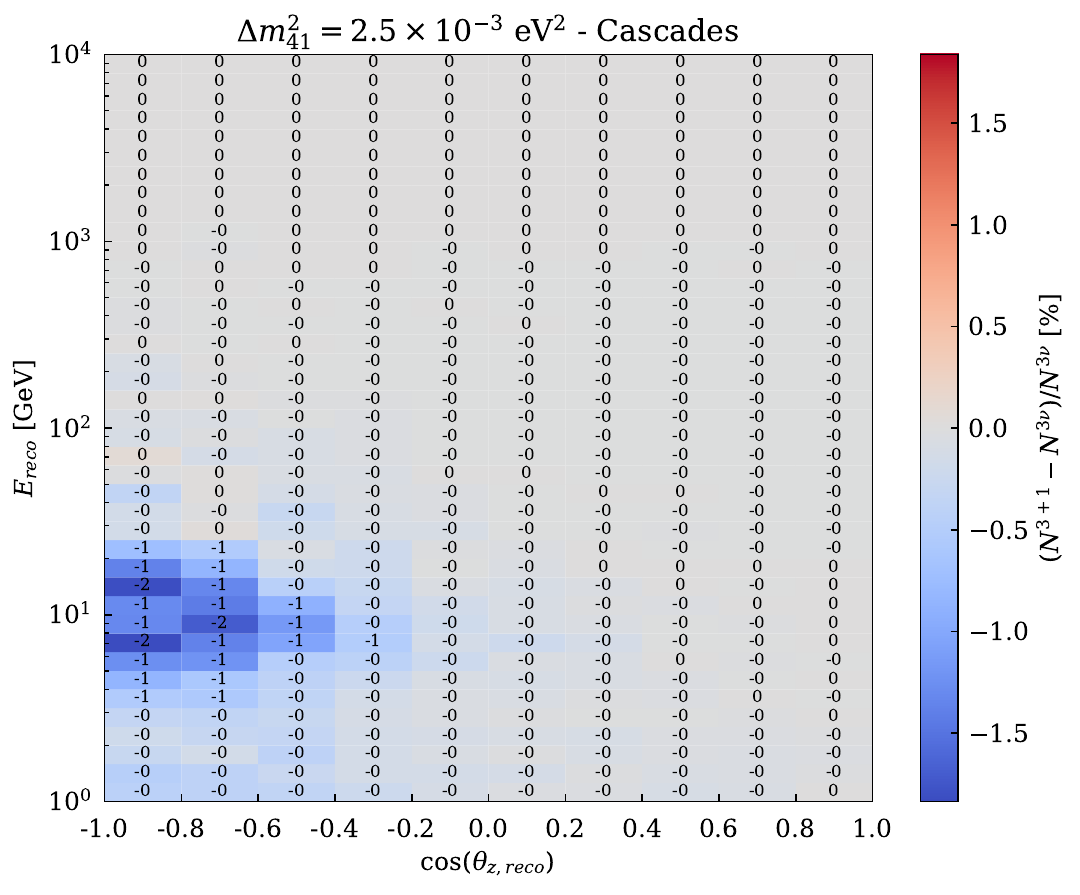}
    \includegraphics[width=0.4\linewidth]{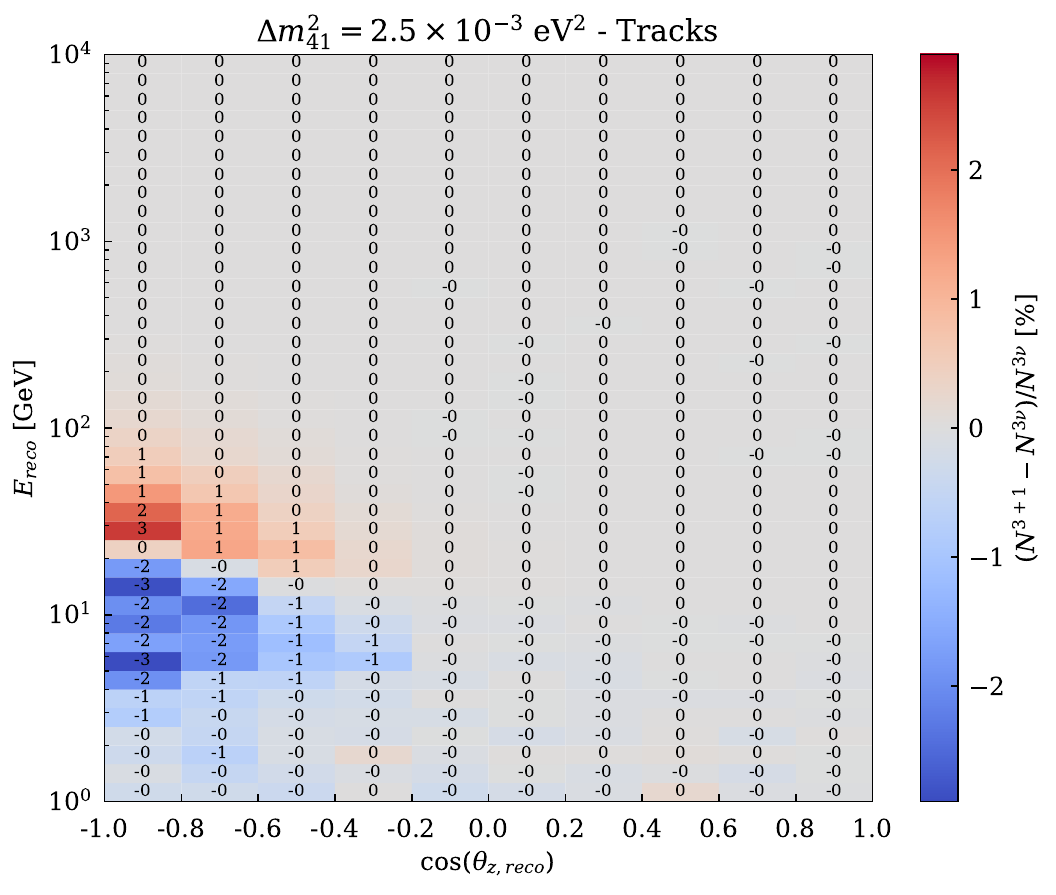}\\
    \includegraphics[width=0.4\linewidth]{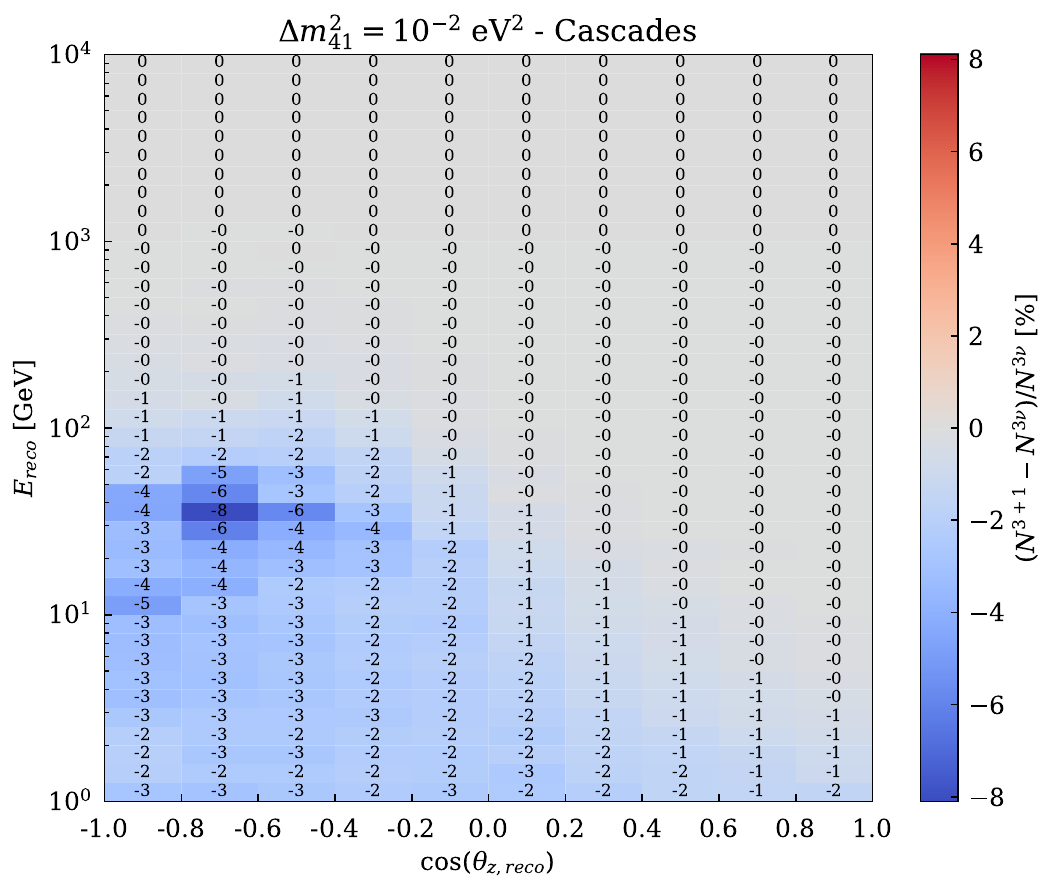}
    \includegraphics[width=0.4\linewidth]{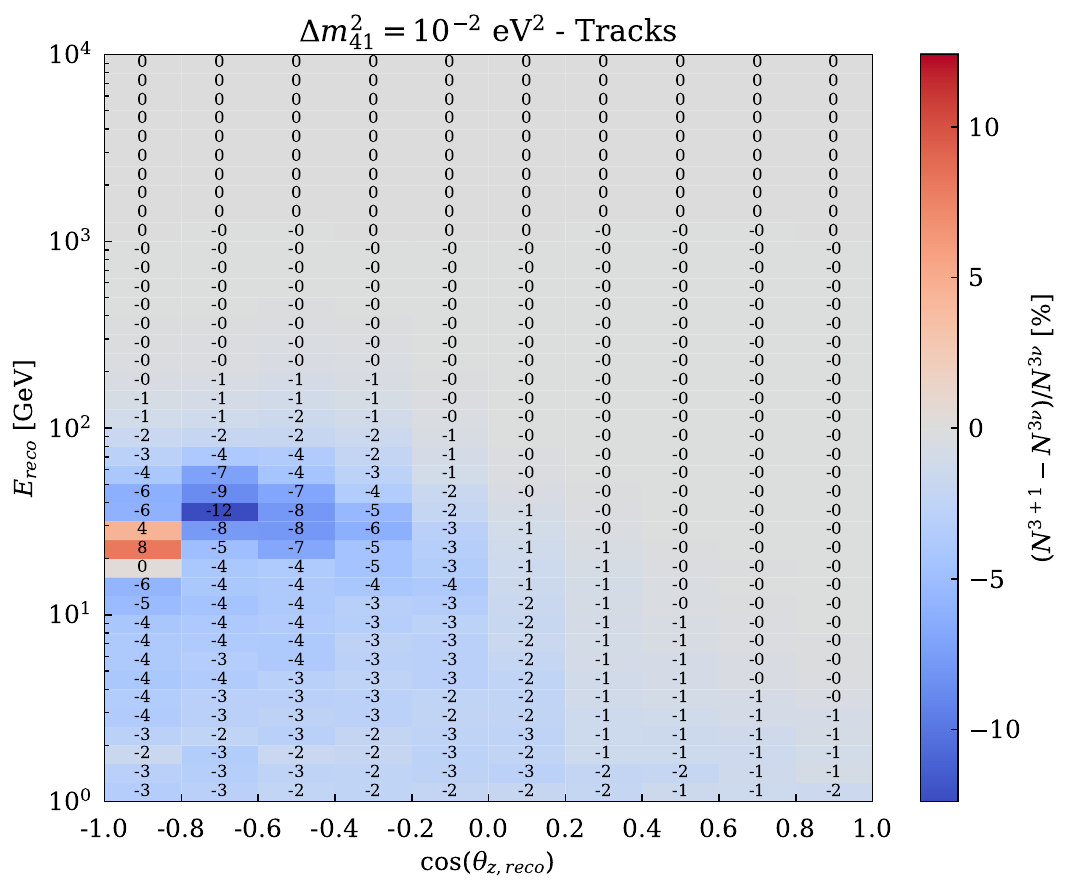}
    \caption{2D distribution of fractional difference in the number of events, $(N^{3+1}-N^{3\nu})/N^{3\nu}\;[\%]$, between the $(3+1)$ and $3\nu$ scenarios, for $\Delta m_{41}^2 = 2.5\times10^{-3}$ and $ 10^{-2}$ eV$^2$.}
    \label{fig:dist2D_1}
\end{figure}

\begin{figure}[ht]
    \centering
    \includegraphics[width=0.4\linewidth]{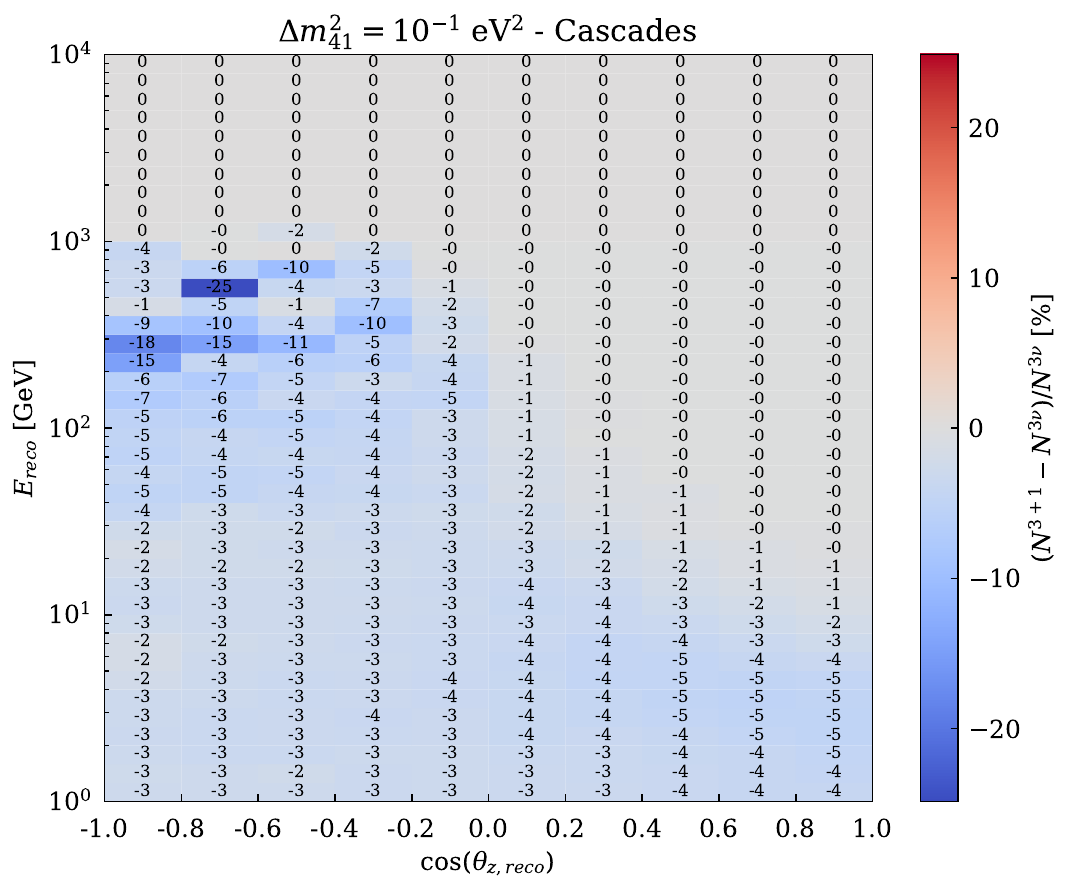}
    \includegraphics[width=0.4\linewidth]{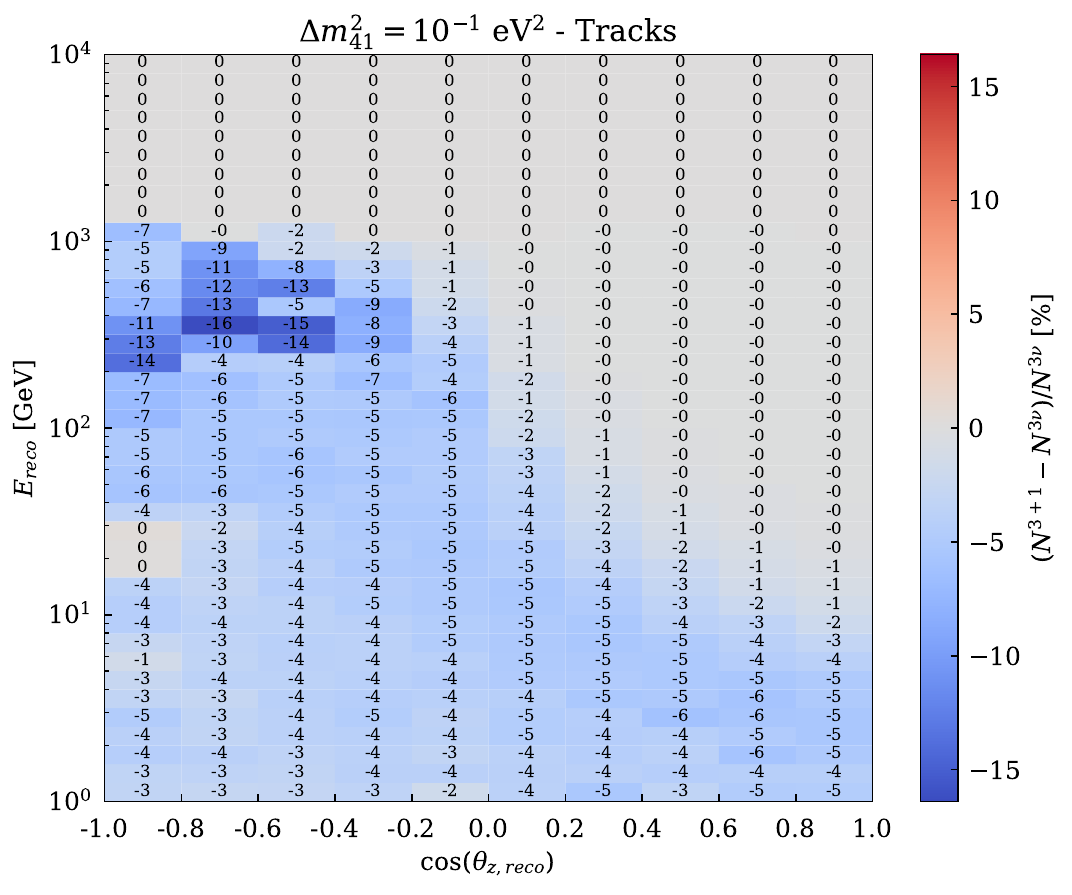}\\
    \includegraphics[width=0.4\linewidth]{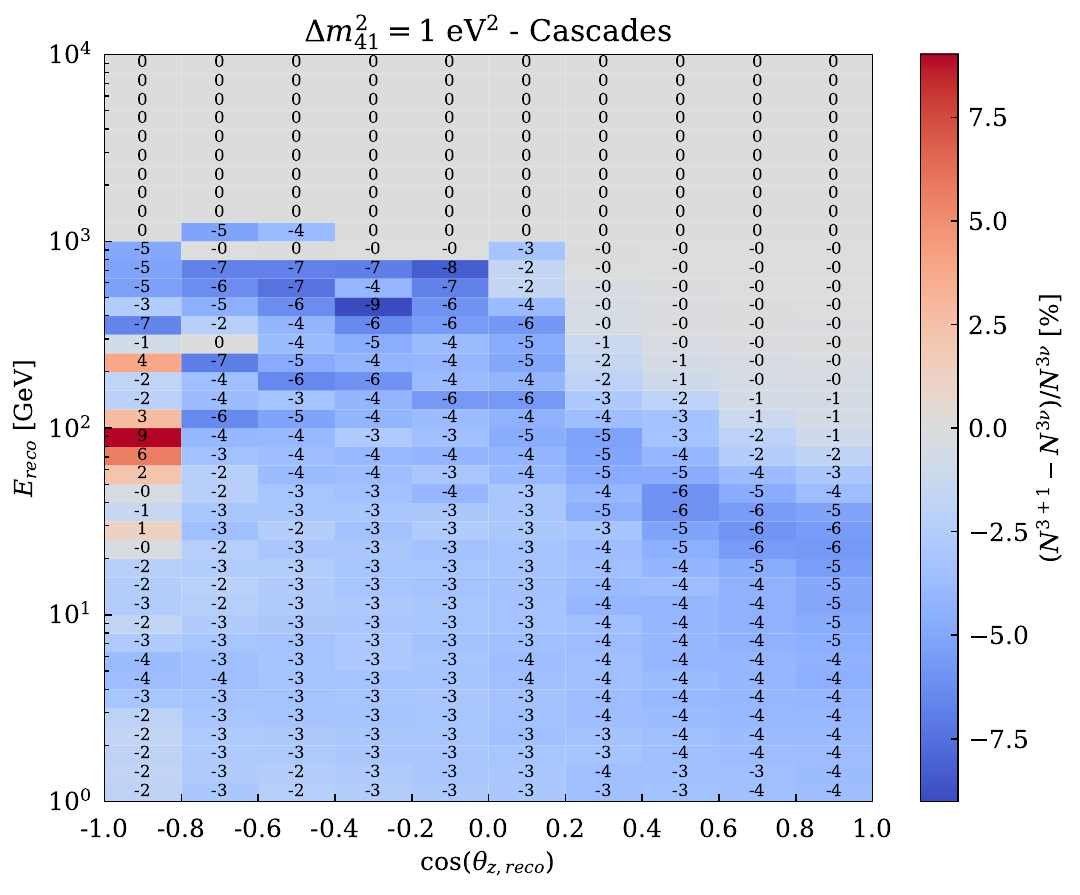}
    \includegraphics[width=0.4\linewidth]{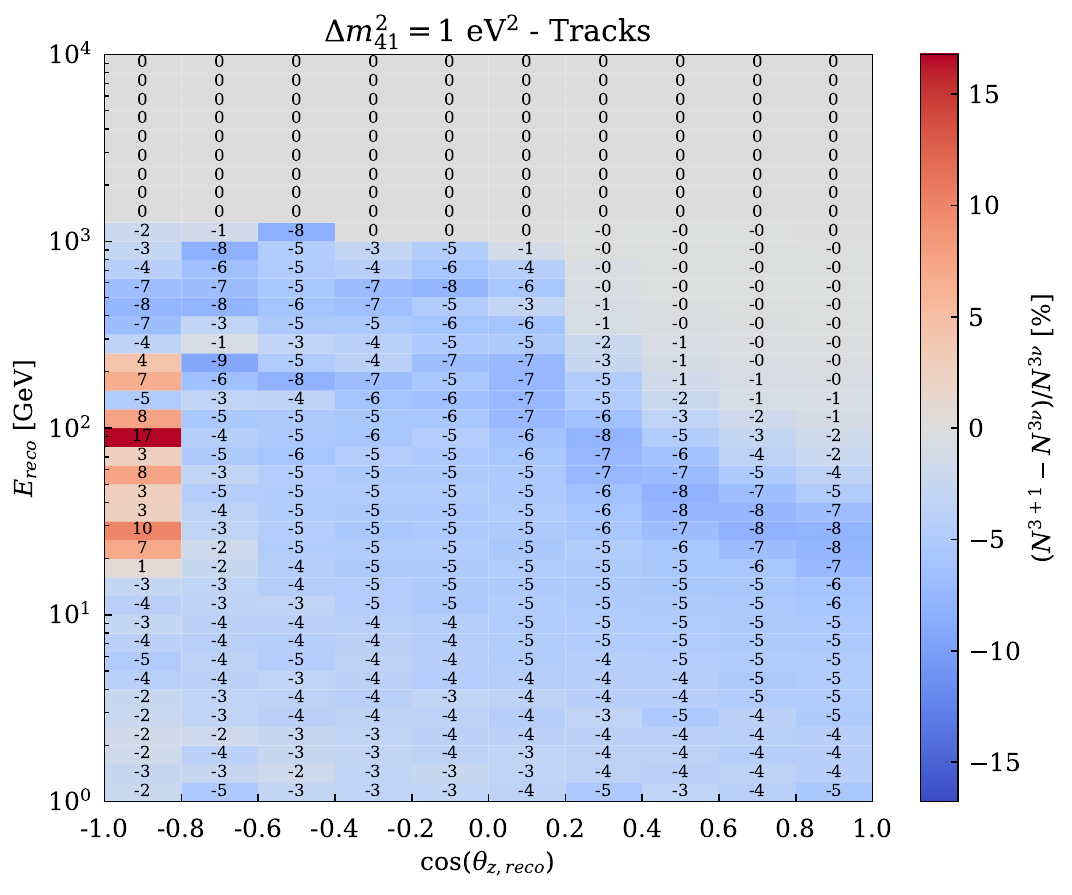}
    \caption{The same as Fig.~\ref{fig:dist2D_1}, but for $\Delta m_{41}^2 = 10^{-1}$ and 1 eV$^2$.}
    \label{fig:dist2D_2}
\end{figure}

\end{document}

%% file: Packages.tex
\usepackage{CJKutf8}

\usepackage{tikz}
\usepackage{adjustbox}
\usepackage{amsmath}
\usepackage{graphicx}
\usepackage{fullpage}
\usepackage[colorlinks=true,allcolors=blue]{hyperref}
\usepackage[capitalise]{cleveref}
\usepackage[utf8]{inputenc}
\usepackage{siunitx}
\usepackage{array}
\input{units}

\usepackage{pifont}
\usepackage{enumitem}
\usepackage{url}
\usepackage{rotating}
\usepackage{epsfig,graphics,rotate}
\usepackage{flushend}
\usepackage{bm}
\usepackage{amssymb}
\usepackage{amsmath}
\usepackage{amsfonts}
\usepackage{gensymb}
\usepackage{booktabs}

\usepackage{microtype}
\usepackage{multirow, makecell}
\usepackage{lipsum}

\usepackage{xspace}
\usepackage{comment}
\usepackage{floatrow}
\usepackage{ragged2e}
\usepackage{siunitx}

\usepackage{nicefrac, xfrac}
\usepackage{mathtools} 
\usepackage{relsize}   

\usepackage{float}

%% file: units.tex
\DeclareSIUnit \s {\second}
\DeclareSIUnit \ns {\nano\second}
\DeclareSIUnit \mus {\micro\second}
\DeclareSIUnit \ms {\milli\second}
\DeclareSIUnit \MB {\mega\byte}
\DeclareSIUnit \GB {\giga\byte}
\DeclareSIUnit \TB {\tera\byte}
\DeclareSIUnit \PB {\peta\byte}
\DeclareSIUnit \Mbps {\mega\bit/\s}
\DeclareSIUnit \Gbps {\giga\bit/\s}
\DeclareSIUnit \Tbps {\tera\bit/\s}
\DeclareSIUnit \Pbps {\peta\bit/\s}
\DeclareSIUnit \kton {\kilo\tonne} 
\DeclareSIUnit \kt {\kilo\tonne}
\DeclareSIUnit \kty {\kilo\tonne-\year}
\DeclareSIUnit \Mt {\mega\tonne}
\DeclareSIUnit \eV {\electronvolt}
\DeclareSIUnit \keV {\kilo\electronvolt}
\DeclareSIUnit \MeV {\mega\electronvolt}
\DeclareSIUnit \GeV {\giga\electronvolt}
\DeclareSIUnit \TeV {\tera\electronvolt}
\DeclareSIUnit \PeV {\peta\electronvolt}
\DeclareSIUnit \EeV {\exa\electronvolt}
\DeclareSIUnit \sr {sr}
\DeclareSIUnit \m {\meter}
\DeclareSIUnit \cm {\centi\meter}
\DeclareSIUnit \nm {\nano\meter}
\DeclareSIUnit \in {\inchcommand}
\DeclareSIUnit \km {\kilo\meter}
\DeclareSIUnit \kV {\kilo\volt}
\DeclareSIUnit \kW {\kilo\watt}
\DeclareSIUnit \MW {\mega\watt}
\DeclareSIUnit \MHz {\mega\hertz}
\DeclareSIUnit \mrad {\milli\radian}
\DeclareSIUnit \year {years}
\DeclareSIUnit \POT {POT}
\DeclareSIUnit \sig {$\sigma$}
\DeclareSIUnit\parsec{pc}
\DeclareSIUnit\lightyear{ly}
\DeclareSIUnit\foot{ft}
\DeclareSIUnit\ft{ft}
\DeclareSIUnit \ppb{ppb}
\DeclareSIUnit \ppt{ppt}
\DeclareSIUnit \samples{S}
\DeclareSIUnit \pe{PE}
\DeclareSIUnit \GeVmwe{GeV/mwe}
\DeclareSIUnit \mwe{mwe}

\newcommand{\enu}{\E_\enu}

%% file: Commands.tex
\definecolor{lime}{HTML}{A6CE39}
\DeclareRobustCommand{\orcidicon}{
	\begin{tikzpicture}
	\draw[lime, fill=lime] (0,0) 
	circle [radius=0.16] 
	node[white] {{\fontfamily{qag}\selectfont \tiny ID}};
	\draw[white, fill=white] (-0.0665,0.095) 
	circle [radius=0.005];
	\end{tikzpicture}
	\hspace{-2mm}
}

\foreach \x in {A, ..., Z}{\expandafter\xdef\csname orcid\x\endcsname{\noexpand\href{https://orcid.org/\csname orcidauthor\x\endcsname}
			{\noexpand\orcidicon}}
   }
